%% file: main.tex
\documentclass[9pt,twocolumn,twoside]{osajnl}

\journal{jocn} 

\setboolean{shortarticle}{false}
\title{Graph Transformers and Stabilized Reinforcement Learning for Large-Scale Dynamic Routing Modulation and Spectrum Allocation in Elastic Optical Networks}

\author[1*]{Michael Doherty}
\author[2]{Alejandra Beghelli}
\author[1]{Laura Toni}

\affil[1]{University College London, Torrington Place, London WC1E 7JE, United Kingdom}
\affil[2]{British Telecom, 1 Braham Street, London E1 8EE, United Kingdom}

\affil[*]{Corresponding author: michael.doherty.21@ucl.ac.uk}

\begin{abstract}
Reinforcement learning (RL) has been widely applied to dynamic routing, modulation and spectrum assignment (RMSA) in optical networks, yet no prior work has trained a transformer model for this task. We attribute this to the high data and compute requirements of transformers and potential training instabilities with RL.
We address this gap by combining recent advances from the machine learning literature (rotary positional encodings for graph-structured data, off-policy invalid action masking, and valid mass regularization) with GPU-accelerated simulation to achieve, for the first time, stable RL training of a transformer for dynamic RMSA. We demonstrate, through systematic benchmarking against previous RL methods and heuristic algorithms, that ours is the first RL method to exceed all benchmarks, increasing the supportable traffic load by up to 13\%.
To demonstrate the scalability of our approach, we train on real network topologies from the TopologyBench database up to 143 nodes and 362 links, with 320 x 12.5\,GHz frequency slot units per link, and 100\,Gbps traffic requests. To our knowledge, these are the largest dynamic RMSA problems to which RL has been applied. We find up to 4\% increased traffic load can be supported at low blocking probability (<0.1\%) with our method compared to the best available benchmark algorithm. We present an ablation study of the components of our training algorithm, the dynamics of the loss function during training, and analyze the allocation decisions of the trained models.
We make all code used to produce this paper openly available for reproduction and future benchmarking: \url{https://github.com/micdoh/XLRON}.
\end{abstract}

\setboolean{displaycopyright}{false} % Do not include copyright or licensing information in submission.

\definecolor{tealrow}{HTML}{E1F6F2}

\begin{document}

\maketitle

\input{PAPER_SECTIONS/1_introduction}

\input{PAPER_SECTIONS/2_transformer_rl}

\input{PAPER_SECTIONS/3_comparison}

\input{PAPER_SECTIONS/4_large_scale}

\input{PAPER_SECTIONS/5_conclusion}

\section*{Acknowledgments}
This work was supported by the Engineering and Physical Sciences Research Council (EPSRC) grant EP/S022139/1 - the Centre for Doctoral Training in Connected Electronic and Photonic Systems - and EPSRC Programme Grant TRANSNET (EP/R035342/1)

% Bibliography
\bibliography{references.bib}

\end{document}

%% file: PAPER_SECTIONS/1_introduction.tex
\section{Introduction}
\label{sec:introduction}

Dynamic routing, modulation and spectrum allocation (RMSA) is a fundamental problem in elastic optical networks (EONs), where incoming traffic requests must be assigned a route, modulation format, and contiguous block of spectrum in real time. Efficient RMSA directly determines network capacity. In practice, operators rely on heuristic algorithms such as $k$-shortest path routing with first-fit spectrum assignment (KSP-FF), which are simple but sub-optimal. Exact optimization via integer linear programming does not scale to realistic network sizes \cite{jaumard_decomposition_2023} or the time constraints of dynamic operation.

Reinforcement learning (RL) offers an appealing middle ground: agents can learn allocation policies from interaction with a simulated network, making decisions in microseconds at inference time while potentially discovering strategies that surpass hand-crafted heuristics. However, a systematic evaluation of nearly 100 papers applying RL to dynamic resource allocation in optical networks \cite{doherty_reinforcement_2025} revealed that many proposed methods were outperformed by simple heuristic baselines that the original studies did not consider.

%Many of these approaches were unable to consider a sufficiently large number of candidate paths, or neglected spectrum allocation entirely by defaulting to first-fit, therefore did not scale to the complexity of realistic problem settings.

One limitation of previous methods has been the choice of model architecture. Transformers have achieved state-of-the-art performance across many domains, owing to their expressive self-attention mechanism and ability to learn representations from sequential and structured data \cite{vaswani_attention_2017}, yet have never been applied to the dynamic RMSA problem. %Several works have employed attention-based Pointer Networks \cite{cheng_ptrnet-rsa_2024}, which can attend over variable-length inputs but are limited in expressivity compared to full transformer models. Graph Attention Networks (GATs) \cite{velickovic_graph_2018} offer an alternative attention-based architecture with graph inductive bias, but their message-passing mechanism creates bottlenecks for gradient propagation and limits attention to local neighbourhoods, and the irregular computation patterns of message passing are inefficient on accelerator hardware such as GPUs, which are optimized for dense matrix multiplications.
Although transformers have been applied in a supervised learning context to static resource allocation in optical networks \cite{chen_transformer-pointer_2026}, training a tabula rasa transformer with RL is known to be challenging. Parisotto et al.\ \cite{parisotto_stabilizing_2019} identified instabilities that arise when optimizing transformer parameters with policy gradient objectives (inherent to most RL algorithms), and proposed architectural modifications to mitigate them. 

In this work, we propose additional modifications to stabilize training and present the first successful application of a transformer model trained with RL for dynamic RMSA. We identify four key challenges in applying transformers with RL to this problem and address each systematically:

\begin{enumerate}
    \item \textbf{Data and computational cost.} Transformers require large amounts of training data, and RL requires many computationally-expensive forward passes. We address both with XLRON \cite{doherty_xlron_2023,doherty_xlron_2024}, our GPU-accelerated simulation and training framework that co-locates environment simulation and policy updates on accelerator hardware \cite{hessel_podracer_2021}, to enable large-scale parallelization and training at up to millions of environment transitions per second.

    \item \textbf{Architectural adaptation to graph-structured data.} Standard transformers are permutation equivariant to the input sequence, therefore are unable to distinguish edge or node features without graph-specific modifications. We employ Wavelet-Induced Rotary Encodings (WiRE) \cite{reid_rotary_2026} to inject graph positional information via the Laplacian spectrum of the network topology, and design pooling strategies for the output logits (Section~\ref{subsec:rope}).

    \item \textbf{Training instability under RL objectives.} Policy gradient optimization of transformers can diverge, particularly with invalid action masking. We adopt Pre-LayerNorm \cite{xiong_layer_2020} and introduce off-policy invalid action masking (Section~\ref{subsec:off_policy_iam}) and valid mass stabilization (Section~\ref{subsec:valid_mass_loss}) to prevent collapse while enabling effective feature learning.
\end{enumerate}

The remainder of this paper is organized as follows. Section~\ref{sec:methodology} details our model architecture and training algorithm. Section~\ref{sec:comparison} presents a direct comparison with five prior RL approaches \cite{chen_deeprmsa_2019,tang_heuristic_2022,xu_deep_2022,shimoda_mask_2021,cheng_ptrnet-rsa_2024} and optimized heuristics on four standard topologies, where our agent consistently achieves the highest supported traffic load. Section~\ref{sec:large_scale} evaluates the scalability of our method on real network topologies from the TopologyBench database \cite{matzner_topology_2024}, reaching up to 143 nodes and 362 links - the largest dynamic RMSA instances to which RL has been applied. We conclude with an ablation study isolating the contribution of each training component and an analysis of the allocation patterns learned by the trained models. All code used for this paper is openly available \cite{doherty_xlron_2023}.

%% file: PAPER_SECTIONS/2_transformer_rl.tex
\section{Background and Methodology}
\label{sec:methodology}

Reinforcement learning was first applied to dynamic resource allocation in optical networks in 2003 \cite{garcia_multicast_2003}, with activity accelerating after DeepRMSA \cite{chen_deeprmsa_2019}. Subsequent works explored reward shaping \cite{tang_heuristic_2022}, graph neural networks \cite{xu_deep_2022}, invalid action masking \cite{shimoda_mask_2021,huang_action_2020}, and pointer networks \cite{cheng_ptrnet-rsa_2024}; we review and benchmark these in Section~\ref{sec:comparison}. Despite this progress, no prior work has successfully trained a transformer with RL for dynamic RMSA.

\begin{figure*}[t]
    \centering
    \includegraphics[width=\linewidth]{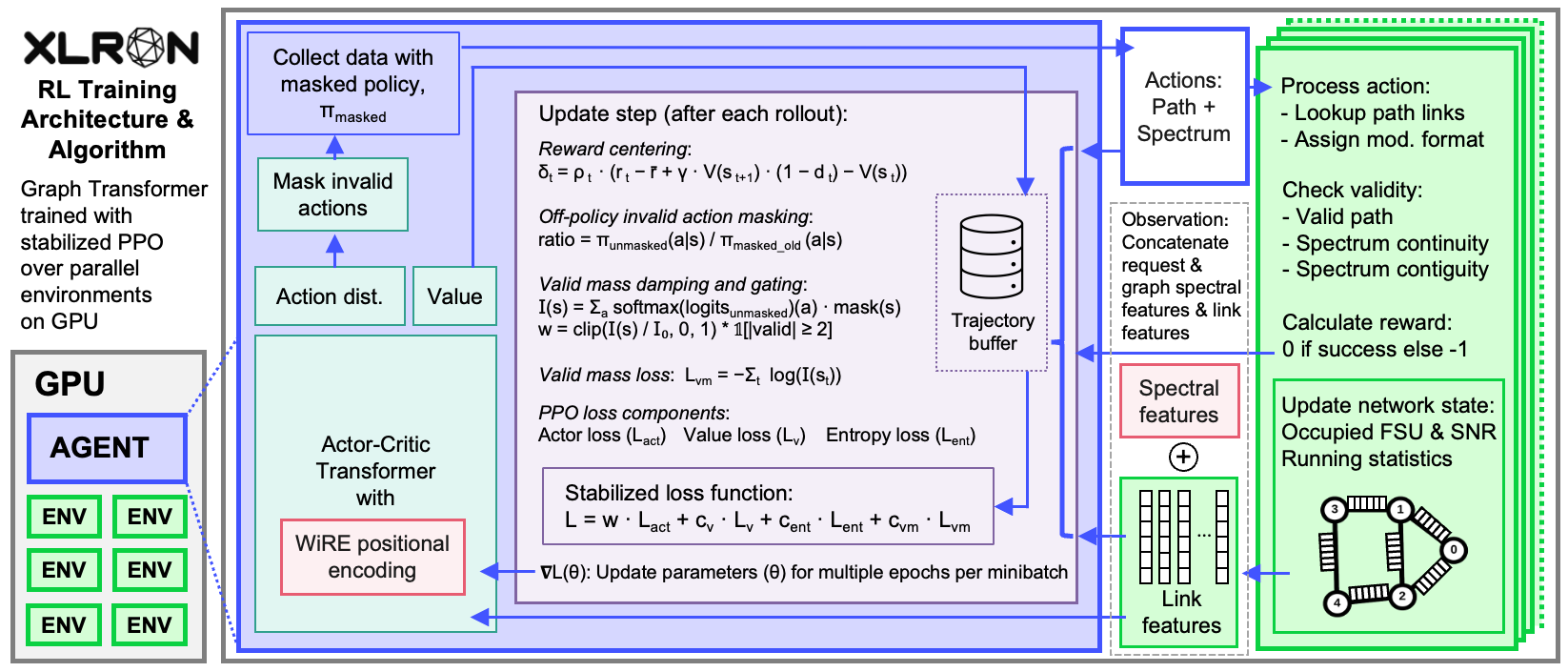}
    \caption{Overview of the XLRON training architecture and algorithm. Parallel environments (green) on GPU generate experience from actions selected by the agent (blue), which comprises a Graph Transformer agent (light blue) trained with stabilized PPO. Key components of the learning algorithm (purple) include off-policy invalid action masking, valid mass stabilization, and WiRE positional encodings (red).}
    \label{fig:training_schematic}
\end{figure*}

Figure~\ref{fig:training_schematic} provides an overview of our training architecture and algorithm. Multiple parallel environments run on GPU (represented in green), each simulating an independent optical network and generating experience concurrently. At each decision step, the environment presents an observation that comprises a series of tokens (many-dimensional vectors). %consisting of a matrix of $N_{\text{links}} \times N_{\text{FSU}}$ elements, where each token is a row of this matrix, representing the spectrum occupancy of a link. 
In our experiments, each token corresponds to a link of the network and contains the the normalized remaining holding time of the connections occupying the link's frequency slots (or zero if unoccupied), though the representation can equally accommodate binary occupancy, estimated SNR, or other per-slot features. We choose holding time for our link features as we found it to give better results than binary occupancy, possibly due to the additional timing information and the regularizing effect of more varied values than binary.

The link features are concatenated with spectral features derived from the topology's Laplacian eigenvectors (Section~\ref{subsec:rope}, Figure~\ref{fig:wire_schematic}) (red boxes in Figure~\ref{fig:training_schematic}) and with features describing the current traffic request (source, destination, required bandwidth). Each link token therefore comprises information about the link's occupancy, its position in the network and the current request. The agent (blue box in Figure~\ref{fig:training_schematic}) comprises the learning algorithm (purple box) and an actor-critic Graph Transformer (Figure~\ref{fig:transformer_schematic}) which processes the tokens and selects an action, subject to invalid action masking. Each action specifies a route and first FSU in which to allocate the request. After each rollout, the collected transitions are used to update the policy via stabilized PPO (Sections~\ref{subsec:rl}--\ref{subsec:valid_mass_loss}). 

The remainder of this section describes each component of the simulation and RL training process outlined in Figure~\ref{fig:training_schematic} in detail: the physical layer assumptions (Section~\ref{subsec:physical_layer}), the RL algorithm and training stabilization (Sections~\ref{subsec:rl}--\ref{subsec:valid_mass_loss}), the transformer architecture and graph positional encodings (Sections~\ref{subsec:transformer}--\ref{subsec:rope}), and the pooling strategies that connect the transformer to the actor-critic outputs (Section~\ref{subsec:pooling}).

\subsection{Physical Layer Model}
\label{subsec:physical_layer}

A growing body of work incorporates quality-of-transmission (QoT) estimation into RL-based resource allocation, computing signal-to-noise ratio (SNR) for each candidate lightpath using analytical models of amplified spontaneous emission noise, nonlinear interference, and inter-channel crosstalk \cite{cheng_ptrnet-rsa_2024,teng_deep-reinforcement-learning-based_2024,teng_drl-assisted_2025,wang_physical_2026}. While these models capture important physical constraints, evaluating per-path SNR at each decision step adds significant computational cost, particularly on large topologies with many candidate and active lightpaths.

An alternative approach, widely adopted in the RMSA literature, assumes worst-case spectrum occupancy to calculate a lower bound per-path SNR value used to determine a maximum transmission distance for each modulation format. These assumptions allow the effects of physical layer impairments to be pre-calculated and captured in the maximum reach, which reduces modulation format selection to a deterministic distance-based lookup. This is the approach we adopt throughout this work. Our goal is to benchmark against previous RL methods under matched problem conditions (Section~\ref{sec:comparison}) and to demonstrate scalability to large topologies (Section~\ref{sec:large_scale}). Since our architecture treats the physical layer model as part of the environment rather than the policy, extending to full SNR-awareness requires no architectural changes and is a natural direction for future work (Section~\ref{sec:conclusion}).

\subsection{Proximal Policy Optimization}
\label{subsec:rl}

We use Proximal Policy Optimization (PPO) \cite{schulman_proximal_2017}, an actor-critic algorithm in which the \emph{actor} outputs a policy $\pi_\theta$ over actions and the \emph{critic} is a value function $V_\phi$ that estimates the expected returns from a state. The advantage $\hat{A}_t$, computed via generalized advantage estimation (GAE) \cite{schulman_high-dimensional_2018}, quantifies how much the observed return exceeds the critic's estimate. PPO constrains each update with a clipped surrogate objective, combining the clipped policy loss, a value function loss $L^{\text{VF}}_t$, and an entropy bonus $H[\pi_\theta]$:

\begin{equation}
\begin{split}
    \mathcal{L}(\theta) = \hat{\mathbb{E}}_t \Big[ &\min\!\left( r_t(\theta) \hat{A}_t,\, \mathrm{clip}(r_t(\theta), 1 - \epsilon, 1 + \epsilon) \hat{A}_t \right) \\
    &- c_V L^{\text{VF}}_t(\theta) + c_{\text{ent}} H[\pi_\theta](s_t) \Big],
\end{split}
\end{equation}

where $r_t(\theta)$ is the probability ratio between the new and old policies, $\hat{A}_t$ is the GAE advantage estimate, $\epsilon$ is the clipping parameter, $L^{\text{VF}}_t$ is the value function loss, $H[\pi_\theta]$ is the entropy of the policy, and $c_V$, $c_{\text{ent}}$ are weighting coefficients.

Applying PPO to train a transformer with invalid action masking introduces specific instabilities that we address in Sections~\ref{subsec:off_policy_iam} and \ref{subsec:valid_mass_loss}.

\subsubsection{Reward Centering}
\label{subsec:reward_centering}

Reward centering \cite{naik_reward_2024} subtracts a running estimate of the mean reward from each observed reward, which has been shown to improve value estimation in continuing (non-episodic) RL tasks and was recently applied to dynamic RMSA by Wang et al.\ \cite{wang_physical_2026}. Dynamic RMSA is naturally a continuing task. However, we find that training with long episodes (10,000 steps or more, sufficient for the blocking rate to reach steady state) produces policies that perform well in both continuing and episodic evaluation settings, while retaining the ability to make effective allocation decisions in a partially-populated or greenfield network at the start of an episode. Episodically-trained policies therefore achieve lower blocking than policies trained on continuing tasks when evaluated over a fixed number of traffic requests, so we use episodic training for the results presented in Sections~\ref{sec:comparison} and~\ref{sec:large_scale}.

\subsection{Off-Policy Invalid Action Masking}
\label{subsec:off_policy_iam}

Invalid action masking (IAM) sets the logits of infeasible actions to $-\infty$ before the softmax, ensuring zero probability on constraint-violating actions. Huang and Onta\~{n}\'{o}n \cite{huang_action_2020} showed that IAM substantially outperforms penalization-based alternatives, and Zabounidis et al.\ \cite{zabounidis_overcoming_2026} proved theoretically that without masking, gradients suppressing invalid actions propagate through shared parameters to suppress valid actions at unvisited states.

Hou et al.\ \cite{hou_exploring_2023} compared on-policy and off-policy forms of IAM. In off-policy IAM, the importance ratio is computed with respect to the \emph{unmasked} policy $\pi_\theta(a|s)$, treating the mask as part of the environment. While on-policy IAM provides exact importance ratio matching, off-policy IAM enables gradient propagation through masked actions, improving feature learning. Based on our experiments and ablation study (Figure~\ref{fig:ablation_blocking}), we adopt off-policy IAM throughout.

\subsection{Valid Mass Stabilization}
\label{subsec:valid_mass_loss}

A failure mode of off-policy IAM arises when the unmasked policy is not incentivized to place more probability mass on valid actions than invalid: the \emph{valid mass} $\mu(s) = \sum_{a \in \mathcal{A}_{\text{valid}}(s)} \pi_\theta(a|s)$ decreases and the importance ratio between the unmasked and masked policies goes towards zero, causing instability.

We address this with three complementary mechanisms. First, a \textbf{valid mass loss} penalizes low valid mass via a log-barrier term:
\begin{equation}
    \mathcal{L}_{\text{VM}}(\theta) = -\hat{\mathbb{E}}_t\!\left[\log \mu_t(\theta)\right],
\end{equation}
where $\mu_t(\theta) = \sum_a \pi_\theta(a|s_t) \cdot m(s_t)$ is the valid mass under the current (unmasked) policy and $m_t$ is the binary action mask. This term is differentiable with respect to the policy logits and encourages the network to maintain probability mass on valid actions throughout training. The coefficient $c_{\text{VM}}$ is optionally annealed over the course of training to reduce its influence on the gradient.

Second, we apply \textbf{per-step loss damping}: for each transition, the actor and entropy loss contributions are scaled by a weight $w_t = \mathrm{clip}(\mu_t / \mu^*, 0, 1)$, where $\mu^*$ is a target valid-mass threshold. A natural calibration point for the target valid mass is the uniform-random baseline: an untrained policy distributing mass uniformly over $|\mathcal{A}|$ actions places $\mu_{\text{uniform}} = |\mathcal{A}_{\text{valid}}|/|\mathcal{A}|$ on valid actions. Setting $\mu^*$ at this baseline ensures that damping activates only when the policy is performing at or below chance at distinguishing valid from invalid actions. For example, in our large-topology experiments with $|\mathcal{A}| \approx$ 200 and heavy network loading where typically only $\sim$1--5\% of actions are valid, we set $\mu^* = 0.05$, corresponding to roughly one to five times the uniform baseline depending on load. Transitions where the policy has little probability mass on valid actions thus receive reduced gradient signal. In combination with the valid mass loss, this is designed to enable off-policy invalid action masking by stabilizing the ratio of the log likelihoods of action selection between the masked and unmasked policies.

Third, we apply \textbf{hard gating}: transitions with fewer than $k_{\min}$ valid actions receive zero weight in the actor and entropy losses. In heavily loaded network states where nearly all spectrum slots are occupied, the agent has no meaningful choice; gating these transitions prevents noise from dominating the gradient.

The full objective becomes:
\begin{equation}
    \mathcal{L}(\theta) = \mathcal{L}_{\text{actor}} + c_1 \mathcal{L}_{\text{VF}} + \mathcal{L}_{\text{entropy}} + c_{\text{VM}} \mathcal{L}_{\text{VM}},
\end{equation}
where $\mathcal{L}_{\text{actor}}$ and $\mathcal{L}_{\text{entropy}}$ incorporate the per-step weights $w_t$. We ablate each component in Section \ref{sec:large_scale}.

\subsection{Transformers}
\label{subsec:transformer}

The transformer \cite{vaswani_attention_2017} uses scaled dot-product attention over query, key, and value projections to enable each input token to attend to all others without the sequential bottleneck of recurrent architectures.

Training transformers with RL is challenging. Parisotto et al.\ \cite{parisotto_stabilizing_2019} showed that standard transformers are often unstable (fail to converge to a useful policy) under policy gradient optimization and proposed gated residual connections (GTrXL). Chen et al.\ \cite{chen_decision_2021} recast RL as conditional sequence modeling (Decision Transformer), though this uses offline data and doesn't take advantage of online simulation. A key choice for stability is the placement of Layer Normalization \cite{ba_layer_2016}: the Pre-LN variant \cite{xiong_layer_2020} applies normalization before each sub-layer, ensuring well-behaved gradients at initialization and eliminating the need for learning rate warm-up. We adopt Pre-LN throughout.

\begin{figure}[ht]
    \centering
    \includegraphics[width=\linewidth]{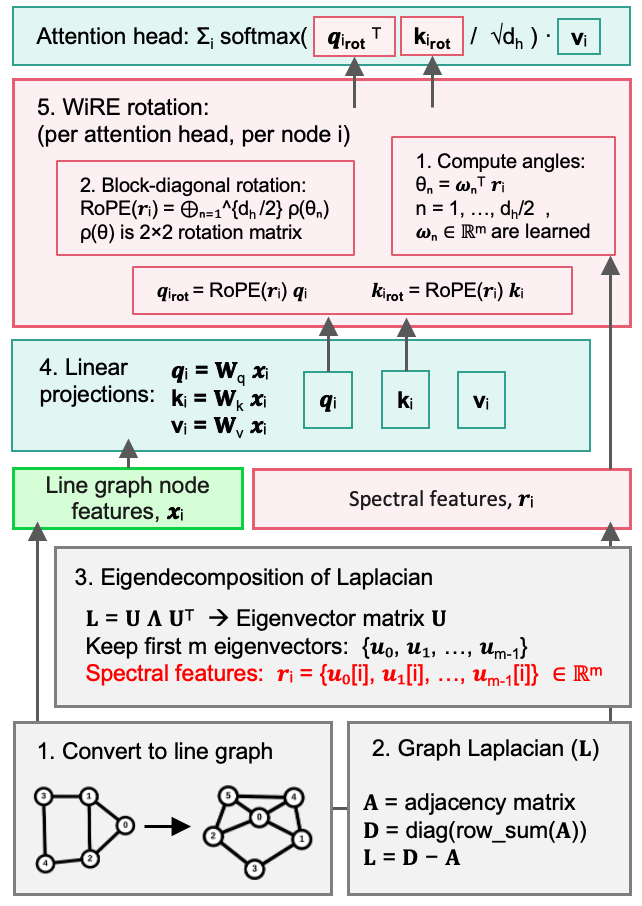}
    \caption{Wavelet-Induced Rotary Encoding (WiRE) for injecting graph positional information into the transformer. The network topology is converted to a line graph, spectral features are extracted from the Laplacian eigenvectors, and rotary position encodings are applied to the query and key vectors in each attention head.}
    \label{fig:wire_schematic}
\end{figure}

\subsection{Graph Transformers}
\label{subsec:graph_transformers}

Applying transformers to graph-structured data requires injecting structural biases. Graph Attention Networks (GATs) \cite{velickovic_graph_2018,brody_how_2022} have been applied to RSA problems \cite{doherty_reinforcement_2025,xiong_graph_2024}, but their message-passing mechanism limits receptive field growth to one hop per layer and is computationally inefficient on GPU hardware.

Graph Transformers overcome these limitations by applying full self-attention over all nodes (or edges), augmented with structural encodings. Graphormer \cite{ying_transformers_2021} demonstrated that the transformer can subsume many GNN variants via centrality, spatial, and edge encodings. GRIT \cite{ma_graph_2023} showed that graph biases can be incorporated without message passing, using learned relative positional encodings. Black et al.\ \cite{black_comparing_2024} proved that absolute and relative positional encodings are equivalent in distinguishing power.

In our architecture, we treat the network links as tokens in the transformer sequence, with each link represented by an embedding of its current spectral occupancy state, as illustrated in the bottom left Figure~\ref{fig:training_schematic}. The critical remaining question is how to encode the graph structure of the network topology into the transformer. We address this in the following section.

\subsection{Rotary Positional Encodings for Graph-Structured Data}
\label{subsec:rope}

Without positional encodings (PE), the transformer is equivariant to input ordering, so PE is essential for injecting structural information. Rotary Position Embedding (RoPE) \cite{su_roformer_2023} has become the dominant PE method, applying rotations to query and key vectors to encode relative position directly in the attention scores.

For graph-structured data, standard sequential PE is inappropriate as there is no canonical ordering of nodes or edges. Recently, Reid et al.\ proposed Wavelet-Induced Rotary Encodings (WiRE) \cite{reid_rotary_2026}, which extends RoPE to graphs via the Laplacian spectrum. WiRE extracts spectral features from the eigenvectors of the graph Laplacian and uses them to parameterize rotary encodings, recovering standard RoPE on regular grids while encoding graph-specific positional information on arbitrary topologies. In this work, we apply WiRE to the link embeddings in our transformer, encoding the position of each link in the network topology via the line graph Laplacian. This allows the model to learn spatial relationships between links without explicit message passing. Figure~\ref{fig:wire_schematic} illustrates the WiRE encoding pipeline from bottom to top. The grey blocks at the bottom of Figure~\ref{fig:wire_schematic} indicate one-time computations of the mapping from the real network topology to its line graph and spectral features that are reused. 

\begin{figure}[ht]
    \centering
    \includegraphics[width=\linewidth]{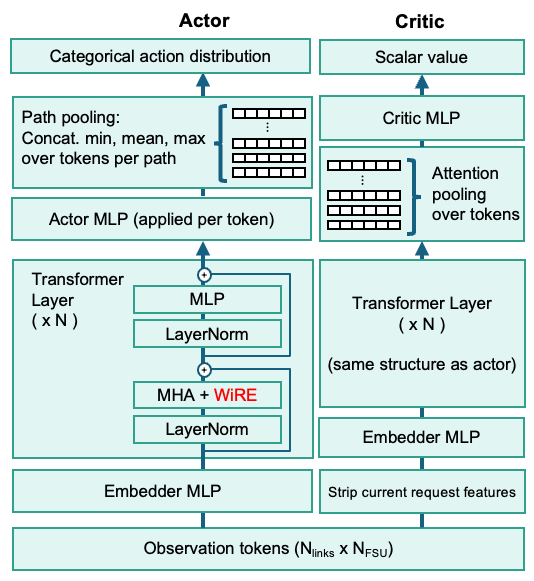}
    \caption{Actor-critic Graph Transformer architecture. The actor uses path pooling (concatenation of min, mean, max over path tokens) to rank candidate lightpaths, while the critic uses learned attention pooling over all tokens to estimate state value.}
    \label{fig:transformer_schematic}
\end{figure}

\subsection{Pooling and Readout}
\label{subsec:pooling}

Converting the transformed output tokens of the transformer into an action or value estimate requires an aggregation or pooling step followed by a readout transformation. The choice of pooling strategy significantly impacts performance \cite{ennadir2025poolmewisely}. In our architecture, the actor and critic benefit from different strategies.

The critic must estimate expected return, which is affected by bottleneck links. Mean pooling dilutes these signals, so we employ single-query attention pooling: a learned vector $\mathbf{q}_v \in \mathbb{R}^d$ attends over the transformer output embeddings $\{\mathbf{h}_l\}$ via
\begin{equation}
    \mathbf{z}_{\text{critic}} = \sum_{l} \alpha_l \, \mathbf{h}_l, \qquad \alpha_l = \mathrm{softmax}\!\left(\mathbf{q}_v^\top \mathbf{h}_l / \sqrt{d}\right),
\end{equation}
introducing only $d$ additional parameters while allowing the critic to learn which links are most informative for value estimation. The pooled feature is transformed by a multi-layer perceptron (MLP) layer to produce the value estimate.

The actor must output a categorical probability distribution over candidate actions. Each action represents a path, each traversing a subset of links $\mathcal{L}_p$, and a FSU. For each path, we compute element-wise statistics (min., mean, and max.) from $\mathcal{L}_p$ and concatenate to form a feature for each path (Equation \ref{eq:pooling}).
\begin{equation}
    \mathbf{z}_p = \left[\min_{l \in \mathcal{L}_p} \mathbf{h}_l^{(\text{act})} \;\Big\|\; \mathrm{mean}_{l \in \mathcal{L}_p} \mathbf{h}_l^{(\text{act})} \;\Big\|\; \max_{l \in \mathcal{L}_p} \mathbf{h}_l^{(\text{act})}\right],
\label{eq:pooling}
\end{equation}

Each path feature is projected by an MLP to action logits. The FSU can optionally be aggregated into sub-bands via \emph{slot aggregation}, reducing the action space dimension; spectrum is then allocated first-fit within the selected sub-band. Figure~\ref{fig:transformer_schematic} illustrates the full actor-critic architecture.

%% file: PAPER_SECTIONS/3_comparison.tex
\section{Comparison of Graph Transformer to Previous RL for Dynamic RMSA}
\label{sec:comparison}

% Another potential problem is the large number of candidate paths as reviewers might think that the larger-K paths are silly unrealistic paths that are too long and unphysical. Butttt if they are too long and unphysical then they will increase blocking a lot and therefore shouldn't be used...? I think that we will need to present some data of the path lengths as K increases for the problems we present to assuage these concerns. I think that the FF-KSP vs Trasnfomer comparisons in the large topology experiments show that the Transformer actually uses shorter paths on avg than the heuristic so as long as we can show that heuristic with large K is still valid then the concern will be resolved.

To evaluate our Graph Transformer architecture, we benchmark against results from five highly cited RL-for-RMSA papers. We used these benchmarks in a previous study \cite{doherty_reinforcement_2025}, which showed that simple heuristics (KSP-FF with paths ordered by hops and sufficient candidate paths) consistently matched or outperformed the published RL solutions. Here, we revisit those same settings and demonstrate that our Graph Transformer agent is the first RL method from the benchmarks to conclusively increase the supportable traffic load compared to the best heuristic algorithms.

\subsection{Selected Benchmark Methods}
\label{subsec:benchmark_methods}

Following our previous benchmarking study, we again compare against five RL methods from the literature, selected for their impact, novelty, and the consistency of their benchmarking practices:

\begin{enumerate}[itemsep=2pt]
    \item \textbf{DeepRMSA} \cite{chen_deeprmsa_2019}: selects from $K$-shortest paths using a fully connected network with first-fit spectrum assignment. Demonstrated 20\% lower service blocking probability (SBP) than KSP-FF on NSFNET and COST239. Its open-source codebase established a de facto benchmark for the field.

    \item \textbf{Reward-RMSA} \cite{tang_heuristic_2022}: extends the DeepRMSA framework with a fragmentation-aware reward function, reporting 32\% lower SBP than heuristics and 55\% lower than DeepRMSA.

    \item \textbf{GCN-RMSA} \cite{xu_deep_2022}: introduces graph convolutional networks with recurrent path aggregation for improved feature extraction, reporting up to 30\% lower SBP than heuristics on NSFNET, COST239, and USNET.

    \item \textbf{MaskRSA} \cite{shimoda_mask_2021}: selects from the full range of available slots across $K$ paths using invalid action masking, reporting over an order of magnitude lower SBP than KSP-FF on NSFNET and JPN48.

    \item \textbf{PtrNet-RSA} \cite{cheng_ptrnet-rsa_2024}: uses Pointer Networks \cite{vinyals_pointer_2015} to construct routes node-by-node rather than selecting from pre-computed paths, combined with invalid action masking for spectrum assignment. Reports over an order of magnitude lower SBP than KSP-FF on NSFNET, COST239, and USNET.
\end{enumerate}

\subsection{Simulation Matching and Methodology}
\label{subsec:simulation_matching}

We recreate each benchmark's problem settings in the XLRON simulation framework \cite{doherty_xlron_2024}, matching topologies, arrival rates, holding times, data-rate distributions, and traffic matrices, under the distance-dependent physical layer assumptions described in Section~\ref{subsec:physical_layer}. Table~\ref{fig:topology_table} summarizes the topologies. We verify fidelity by confirming that KSP-FF with $K$=5 in our framework reproduces the SBP values from the original papers. We also ran the source code for benchmarks and used it to check our simulation is accurate and, where source code was unavailable, we corresponded with the original authors to verify settings.

For each topology and traffic load, we run 10 independent episodes, each consisting of a 3\,000-request warm-up period followed by 10\,000 requests, and report the mean and standard deviation of the SBP. Published RL results are extracted from the original papers. As the best heuristic benchmark, we use KSP-FF with $K$=50 and paths ordered by number of hops, and FF-KSP for JPN48 topology, which were shown in \cite{doherty_reinforcement_2025} to provide the lowest blocking across these problem settings.

% \begin{figure}[h]
%     \centering
%     \includegraphics[width=\linewidth]{IMAGES/topologies/large_topology_table.png}
%     \caption{Summary of network topologies used in the comparison study, ordered by increasing number of nodes.}
%     \label{fig:topology_table_img}
% \end{figure}

\begin{table}[h]
\centering
\small
\setlength{\tabcolsep}{4pt}
\renewcommand{\arraystretch}{1.2}
\begin{tabular}{l c c c c c}
\toprule
\textbf{Topology} & \textbf{Nodes} & \textbf{Dir.\ Links} & \textbf{Avg Deg.} & \textbf{Avg SP} & \textbf{Avg SP} \\
& & & & \textbf{(hops)} & \textbf{(km)} \\
\midrule
\rowcolor{tealrow} NSFNET & 14 & 44 & 1.57 & 2.12 & 2054 \\
COST239 & 11 & 52 & 2.37 & 1.56 & 1810 \\
\rowcolor{tealrow} USNET & 24 & 86 & 1.79 & 2.99 & 2993 \\
JPN48 & 48 & 164 & 1.71 & 5.21 & 1201 \\
\rowcolor{tealrow} USA100 & 100 & 342 & 1.71 & 6.52 & 2684 \\
TATAIND & 143 & 362 & 1.26 & 9.87 & 1972 \\
\bottomrule
\end{tabular}
\caption{Summary of network topologies used in this study, ordered by increasing number of nodes. (Dir. Links) Count of directed single-fibre links in the topology. (Avg Deg.) Average node degree. (SP) Shortest path between node pairs. The larger topologies (USA100 and TATAIND) are used in Section~\ref{sec:large_scale}.}
\label{fig:topology_table}
\end{table}

\subsection{Network Capacity Bounds}
\label{subsec:capacity_estimates}

To contextualize absolute performance, we include two methods to estimate the lower-bound blocking probability obtained by relaxing either the spectrum continuity contraints (cut-sets method) or "no reconfiguration" constraint (resource-prioritized defragmentation). We define the network capacity as the load that can be supported at a target blocking probability, therefore a lower bound blocking estimate provides an upper bound network capacity estimate.

\subsubsection{Cut-set bound}
The cut-set method \cite{hayashi_cost-effective_2023,cruzado_effective_2023,cruzado_capacity-bound_2024} relaxes the spectrum continuity constraint. It identifies the most congested cut-sets in the network and simulates traffic allocation using only one representative link per cut-set traversed by each request, thereby removing the continuity constraint across multi-link paths. We include up to the top 256 most congested cut-sets in each topology to ensure enough cut-sets are considered that blocking is not underestimated.

\subsubsection{Defragmentation bound}
Resource-prioritized defragmentation \cite{doherty_reinforcement_2025} relaxes the no-reconfiguration constraint. Upon each new arrival, all active connections are removed then reallocated in descending order of required FSU using KSP-FF, a technique inspired by allocation methods for static RWA \cite{baroni_routing_1998} and dynamic RWA \cite{beghelli_resource_2006}. By allowing full reconfiguration, fragmentation is eliminated and the resulting SBP presents a bound on what its equivalent non-reconfiguring algorithm can achieve.

\subsection{Transformer Training}
\label{subsec:transformer_training}

For each problem setting, we train a Graph Transformer policy with our custom architecture using the stabilized PPO algorithm, as described in Section~\ref{sec:methodology}. Training used a NVIDIA A100 GPU, with training times from 37m to 1h\,40m depending on the topology and number of steps.

% DeepRMSA NSFNET: 30m, 80M steps, 50K SPS
% DeepRMSA COST239: 45m, 100M steps, 38K SPS
% DeepRMSA USNET: 50m, 50M steps, 20K SPS
% Ptrnet-RSA40 NSFNET: 37m, 100M steps, 47K SPS

All models use a 2-layer transformer with 128-dimensional embeddings, a rollout length of 64, and a slot aggregation factor. The policy and value functions use identical architectures with different pooling mechanisms (Section \ref{subsec:pooling}) and no parameter sharing. We chose the model dimensions to ensure sufficient capacity for feature learning while minimizing memory requirements and training time. The rollout length is sufficient to capture sequential action dependencies without over-relying on value function bootstrapping, while remaining within GPU memory limits. The aggregation factor is used to reduce the total number of possible actions and therefore speed up the learning process, while still maintaining sufficient granularity of spectrum selection for the agent to develop strategies in this dimension. A trained policy and value function contain approximately 600k parameters each and requires 2.5MB on disk. 

The PPO hyperparameters are mostly identical across problem settings: clip $\epsilon$=0.04, discount $\gamma$=0.996, GAE $\lambda$=0.99, cosine learning rate schedule, and a separate lower-learning-rate optimizer for the value function. The valid mass loss coefficient and schedule are held constant ($c_{\text{VM}}$=0.0001--0.001, constant schedule) across all runs. The number of parallel environments ranges from 32 to 256 depending on topology size, and total training timesteps range from 32M to 100M, with larger topologies requiring longer training. The number of attention heads (4 or 8), actor learning rate ($4 \times 10^{-3}$--$6 \times 10^{-3}$), entropy coefficient (0.012--0.0225), and value function coefficient (0.01--0.1) are lightly tuned per problem setting. The traffic load for training in each case was selected from the heuristic data to have 0.1\% to 1\% blocking probability. We reason this produces sufficient blocking signal for the agent to learn from, with multiple blocking events per batch on average, with large relative changes in reward from improved allocations.

In most cases, the trained models reached lower blocking probabilities than the heuristics without extensive hyperparameter tuning, but some required additional effort. All models use a slot aggregation factor of 20, except DeepRMSA NSFNET which uses 100, i.e. the model selects paths only with first-fit spectrum allocation, which enabled the model to learn a policy that slightly improved on the heuristic. Similarly, the JPN48 MaskRSA model uses a two-stage training procedure in which a second run with a reduced learning rate and entropy coefficient fine-tunes the model from the first stage, which was necessary to improve on the heuristic. %The USNET DeepRMSA model is trained with continuous (non-episodic) operation and reward centering, which led to slightly improved blocking compared to episodic training in this case (Section~\ref{subsec:reward_centering}).

\subsection{Results}
\label{subsec:comparison_results}

\begin{figure*}[t]
    \centering
    \includegraphics[width=\linewidth]{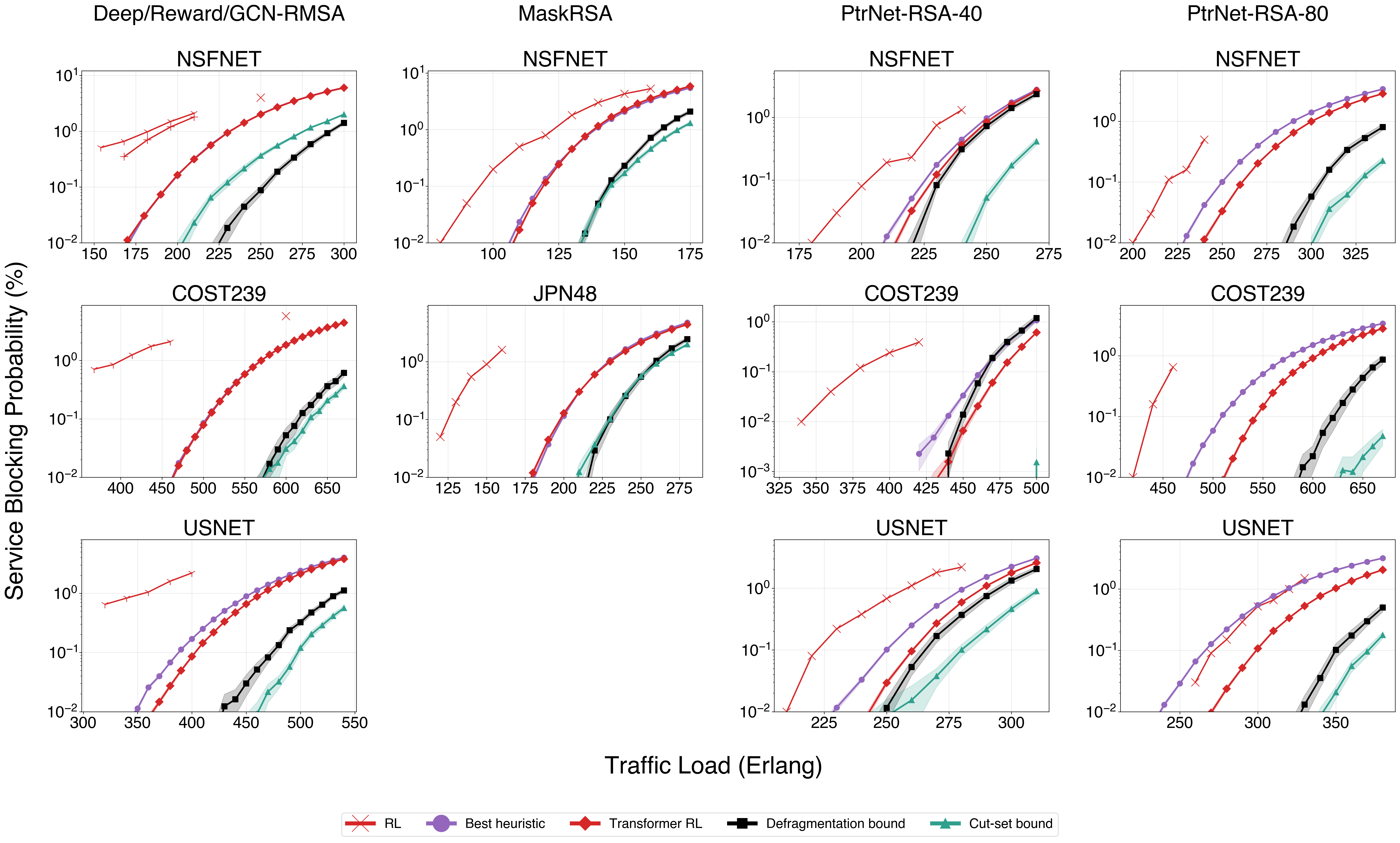}
    \caption{Service blocking probability as a function of traffic load for our method (Transformer RL) compared to previous RL methods (Deep/Reward/GCN-RMSA, MaskRSA, PtrNet-RSA), the best heuristic in each, and upper bound network capacity estimates (defragmentation bound and cut-set bound) across NSFNET, COST239, USNET, and JPN48 topologies. Shaded regions indicate the standard error of the mean across parallel environments. In the first column, DeepRMSA results are shown as a single red $\times$ per plot, RewardRMSA and GCN-RMSA as thin red lines with $+$ and $Y$ markers, respectively.}
    \label{fig:comparison_results}
\end{figure*}

Figure~\ref{fig:comparison_results} presents SBP versus traffic load for our Graph Transformer RL method (red with square markers) alongside the five previous RL methods (thin red lines), the best heuristic (purple), and the two capacity bound estimates (black and teal). Each column is for a common set of simulation settings from the RL method papers detailed in Section~\ref{subsec:benchmark_methods}, with each plot within a column showing results for the labeled topology. The first column shows the results from up to three different previous works per plot (DeepRMSA, RewardRMSA, GCN-RMSA), as each of these works used identical simulation settings per topology therefore have identical heuristic benchmarks.

The capacity bound estimates show differing results. The cut-sets bound tends to be higher capacity (lower blocking) than the defagmentation bound in the cases presented, with the exception of the upper left plot. Both bounds show lower blocking probability each traffic load than the RL methods and heuristic, with the exception of PtrNet-RSA-40 COST239 where our method achieves lower blocking than the defragmentation bound. This is striking as it suggests our method found an allocation strategy that can support more traffic than even if defragmentation operations are permitted. The bounds methods show that overall the allocation methods are close to the upper bounds, with potentially 20\% increased traffic load maybe possible over the heuristic in the most extreme case (PtrNet-RSA-80 USNET).

Looking at the benchmarks results of previous RL methods, they give significantly higher blocking probability than the heuristic. Only for PtrNet-RSA-80 USNET does the RL benchmark slightly improve on the heuristic. This data reasserts our previous findings \cite{doherty_reinforcement_2025} that RL benchmarks are often weaker than a simple heuristic (KSP-FF/FF-KSP) with sufficient candidate paths and optimized path sort criteria.

Our method ("Transformer RL" in Figure~\ref{fig:comparison_results}) improves on the RL benchmarks in all cases. For DeepRMSA and MaskRSA, our method matches or slightly improves on the blocking performance of the best heuristic for the COST239, NSFNET and JPN48 topologies, with almost complete overlap of the curves in some cases. For USNET, our method achieves approximately 4\% higher traffic load at 0.1\% blocking then the heuristic.

The greatest improvements are seen in the PtrNet-RSA cases (third and fourth columns). 
In all cases our method supports higher traffic loads, with the greatest improvement in on PtrNet-RSA-80 USNET, with 13\% higher supported traffic load than the heuristic. In the PtrNet-RSA-40 COST239 case, our method even improves on the defragmentation bound (which offers limited improvement over the KSP-FF heuristic in this case) and suggests that a superior routing policy has been discovered.

Overall, our method achieves the lowest blocking across all four topologies, with the largest improvement on USNET under PtrNet-RSA-80 conditions (up to 13\% higher supported traffic load). The gap between our Graph Transformer and the bound estimates is substantially narrower than that of any competing method, though it suggests some further gains may be possible through improved routing and spectrum management.

%% file: PAPER_SECTIONS/4_large_scale.tex
\section{Large-Scale Experiments}
\label{sec:large_scale}

\begin{figure}[htbp]
    \centering
    \includegraphics[width=\linewidth]{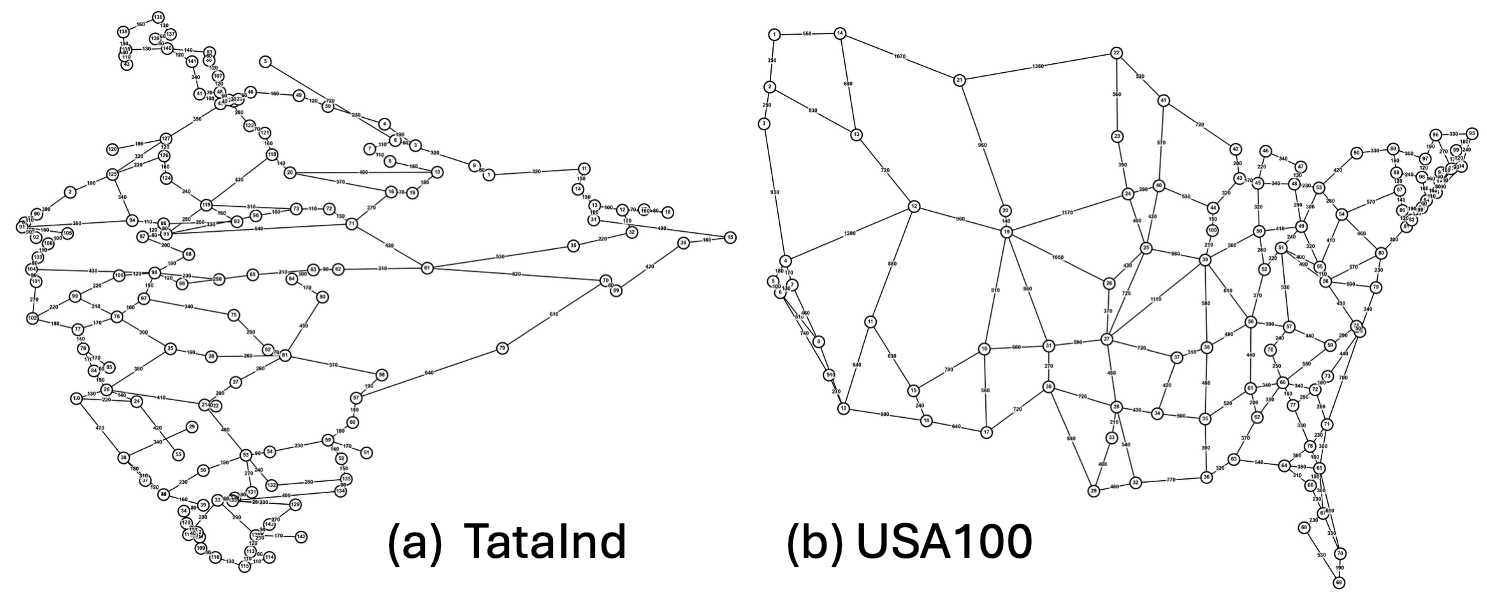}
    \caption{The (a) TataInd and (b) USA100 network topologies used in the large-scale experiments. Node labels indicate node IDs and edge labels indicate link lengths in km.}
    \label{fig:topologies}
\end{figure}

We now evaluate scalability on topologies substantially larger than any previously attempted in the dynamic RMSA literature. We select TataInd (143 nodes, 362 links) and USA100 (100 nodes, 342 links) from the TopologyBench database \cite{matzner_topology_2024} (Figure~\ref{fig:topologies}), both with 320 $\times$ 12.5\,GHz FSU per link, 100\,Gbps requests, and distance-dependent modulation formats. To our knowledge, these are the largest dynamic RMSA instances to which RL has been applied.

\subsection{Heuristic Benchmarks}
\label{subsec:heuristic_benchmarks}

For the TataInd and USA100 topologies, we systematically determine the strongest heuristic benchmark. Previous work \cite{doherty_reinforcement_2025} showed that FF-KSP and KSP-FF heuristics provided the lowest blocking probability compared to five other other heuristics, and that blocking decreases as the number of candidate paths $K$ increases until reaching an asymptotic minimum. We use these insights and compare the blocking performance of both heuristics for $K$ from 10 to 100. We select a traffic load that produces between 0.1\% and 1\% blocking (450 Erlang for TataInd and 620 Erlang for USA100) and run a simulation for 100k requests for each $K$ value, finding that FF-KSP with K=90 and K=70 give lowest blocking. 

We note these K-values are much higher than any explored in previous work. In order to justify considering a high number of candidate paths, we first point out that goal here is to establish the strongest possible benchmark against which to evaluate the Transformer agent. We therefore explore the effect of increased candidate paths K. The topologies chosen in this study are able to support a higher number of candidate paths compared to previous works because they are much larger (in terms of node and edge count) than any previously considered topologies in studies of RL for RMSA. 
Additionally, the average shortest path between any node pair on the topologies are 6.52 (USA100) and 9.87 (TataInd). These many-hop paths permit a large number of link permutations (resulting in many viable candidate paths) without excessively increasing hop count or total path length in km. 

We substantiate our choice of heuristic and assertion that K=90 and K=70 are acceptable values by recording the statistics of the paths selected by the KSP-FF and FF-KSP heuristics each over a 100k request episode and analyze them. FF-KSP achieves substantially lower blocking than KSP-FF on both topologies: 0.72\% versus 3.53\% SBP on USA100 ($K$=70, 620~Erlang) and 0.90\% versus 3.61\% on TataInd ($K$=90, 450~Erlang). This comes at the cost of longer paths; FF-KSP uses paths that are 400--500\,km longer on average than KSP-FF (mean 3,655\,km versus 3,235\,km on USA100; 2,778\,km versus 2,447\,km on TataInd) and routes only 28--37\% of traffic on the shortest path, compared to 77--81\% for KSP-FF. The longer paths require lower-order modulation formats that consume more spectrum per channel, yet the more-diverse routing more than compensates for this penalty. Importantly, the paths used by FF-KSP are not unusably long: over 99.9\% of accepted paths are under 7,500\,km, with mean hop counts of 7.8 (USA100) and 12.0 (TataInd). These mean path lengths in hops are approximately two hops longer than the average shortest path between any two nodes, which is not an excessive increase. Under more stringent physical layer constraints (e.g.\ nonlinear impairments, transceiver noise, insertion loss at ROADMs and other network components), the optimal heuristic settings would likely differ, but our goal here is to establish the strongest possible benchmarks for this problem setting. We therefore adopt FF-KSP with $K$=70 and $K$=90 as the heuristic benchmark for USA100 and TataInd respectively.

\subsection{Training Configuration}
\label{subsec:training_config}

We train Graph Transformer policies for each topology using the stabilized PPO algorithm and architecture described in Section~\ref{sec:methodology}. Training is performed entirely on a single NVIDIA H100 GPU using the XLRON framework, with 12 parallel environments generating experience concurrently. Each training run uses 40M traffic request steps (approximately 52,000 policy updates with a rollout length of 64 and batch size of 768 transitions). Episodes consist of 25,000 requests each at the training load (450~Erlang for TataInd, 620~Erlang for USA100), sufficient for the network to reach steady-state occupancy. The spectrum is divided into sub-bands of 80~FSU via slot aggregation, yielding an action space of $K \times 4$ ($K$=90 for TataInd, $K$=70 for USA100). Both topologies share the same transformer architecture (2 layers, 8 attention heads, embedding dimension 128) and most PPO hyperparameters; the two topologies differ only in entropy coefficient and valid mass loss coefficient, as summarised in Table~\ref{table:hyperparams}. A separate optimizer with a lower learning rate is used for the value function.

Thanks to the compilation and parallelism offered by the XLRON framework, training completes in a short enough time to enable multiple training runs, despite the scale of these topologies: the USA100 policy trains in 4\,h\,21\,min (2,640 steps/s) and TataInd in 5\,h\,15\,min (2,190 steps/s), with compilation completing in 123\,s and 253\,s respectively. Both runs consume 74\,GB of GPU memory on a single H100.

\begin{table}[ht]
\centering
\small
\setlength{\tabcolsep}{6pt}
\renewcommand{\arraystretch}{1.2}
\begin{tabular}{l c c}
\toprule
\textbf{Parameter} & \textbf{TataInd} & \textbf{USA100} \\
\midrule
\multicolumn{3}{c}{\textit{Environment}} \\
\midrule
\rowcolor{tealrow} Candidate paths $K$ & 90 & 70 \\
Link resources (FSU) & 320 & 320 \\
\rowcolor{tealrow} Training load (Erlang) & 450 & 620 \\
Episode length (requests) & 25,000 & 25,000 \\
\midrule
\multicolumn{3}{c}{\textit{Transformer architecture}} \\
\midrule
\rowcolor{tealrow} Layers & 2 & 2 \\
Attention heads & 8 & 8 \\
\rowcolor{tealrow} Embedding dimension & 128 & 128 \\
FSU aggregation factor & 80 & 80 \\
\rowcolor{tealrow} Actor pooling & min/mean/max & min/mean/max \\
Critic pooling & attention & attention \\
\midrule
\multicolumn{3}{c}{\textit{PPO and training}} \\
\midrule
\rowcolor{tealrow} Total timesteps & 40M & 40M \\
Parallel environments & 12 & 12 \\
\rowcolor{tealrow} Rollout length & 64 & 64 \\
Learning rate (actor) & $1.5 \times 10^{-3}$ & $1.5 \times 10^{-3}$ \\
\rowcolor{tealrow} Learning rate (critic) & $5 \times 10^{-5}$ & $5 \times 10^{-5}$ \\
LR schedule & cosine & cosine \\
\rowcolor{tealrow} Clip $\epsilon$ & 0.04 & 0.04 \\
Discount $\gamma$ & 0.996 & 0.996 \\
\rowcolor{tealrow} GAE $\lambda$ & 0.99 & 0.99 \\
Value function coeff.\ $c_V$ & 0.1 & 0.1 \\
\rowcolor{tealrow} Entropy coeff.\ $c_{\text{ent}}$ & 0.015 & 0.01 \\
Entropy schedule & cosine & cosine \\
\rowcolor{tealrow} Valid mass loss coeff.\ $c_{\text{VM}}$ & 0.001 & 0.002 \\
VML schedule & linear & linear \\
\rowcolor{tealrow} VML end fraction & 0.5 & 0.5 \\
\bottomrule
\end{tabular}
\caption{Hyperparameters for training on TataInd and USA100. Parameters that differ between topologies are the number of candidate paths, training load, entropy coefficient, and valid mass loss (VML) coefficient.}
\label{table:hyperparams}
\end{table}

\subsection{Ablation Study}
\label{subsec:ablation}

\begin{figure*}[t]
    \centering
    \includegraphics[width=\linewidth]{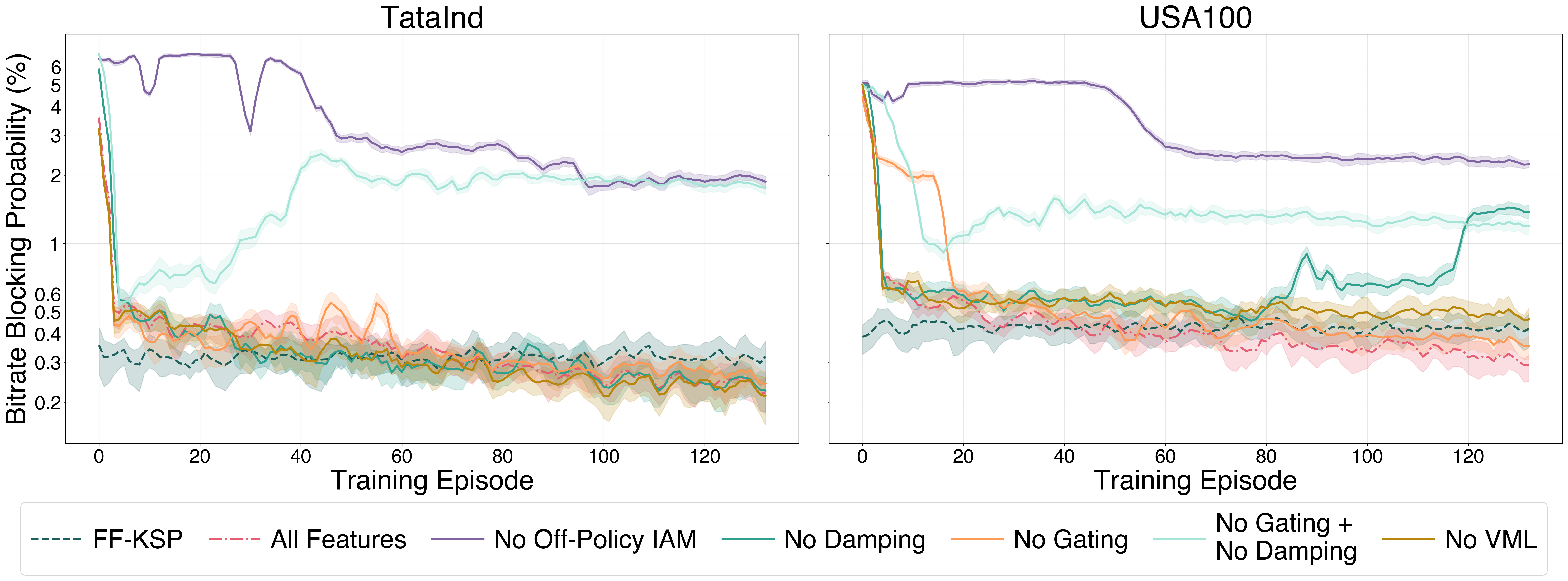}
    \caption{Ablation study of key training components on TataInd and USA100 topologies. Each curve removes one component from the full model (All Features). The FF-KSP heuristic benchmark is shown for reference. Shaded regions indicate the upper and lower interquartile range across parallel environments.}
    \label{fig:ablation_blocking}
\end{figure*}

To understand the importance of each component of our training algorithm to the outcome of training, we perform an ablation study in which we remove one component at a time and retrain. Figure~\ref{fig:ablation_blocking} shows the blocking probability over the course of training for each ablated variant alongside the full algorithm (``All Features'') and the FF-KSP heuristic benchmark.

% TODO - update figure to move legend underneath

The most critical component is off-policy invalid action masking: removing it (``No Off-Policy IAM'', reverting to on-policy masking) prevents convergence on both topologies, with blocking remaining several times higher than the heuristic throughout training. This confirms the finding of Hou et al.\ \cite{hou_exploring_2023} that the choice of masking strategy has a large effect on learned performance, and suggests that off-policy IAM is required for the transformer to learn effective feature representations.

Per-step loss damping is the second most important component. Without damping (``No Damping''), training on USA100 exhibits a dramatic instability: blocking drops initially but then collapses to well above the heuristic level. On TataInd, removing damping leads to high-variance training that converges to a substantially worse policy. Combining the removal of both damping and gating (``No Gating + No Damping'') produces similarly unstable behaviour, suggesting that damping is the primary stabilizer among the valid mass mechanisms.

Removing the valid mass loss entirely (``No VML'') has differing effects in each case: on USA100 it causes a late training collapse similar to the no-damping case, while on TataInd the effect is more modest. Removing gating alone (``No Gating'') has the smallest effect on both topologies, with the ablated variant converging close to the full model. Overall, it is clear that some valid mass regularization is required to prevent the policy diverging and the ratio between the masked behavior policy and unmasked target policy collapsing to zero.

\subsection{Training Dynamics}
\label{subsec:training_dynamics}

\begin{figure*}[t]
    \centering
    \includegraphics[width=\linewidth]{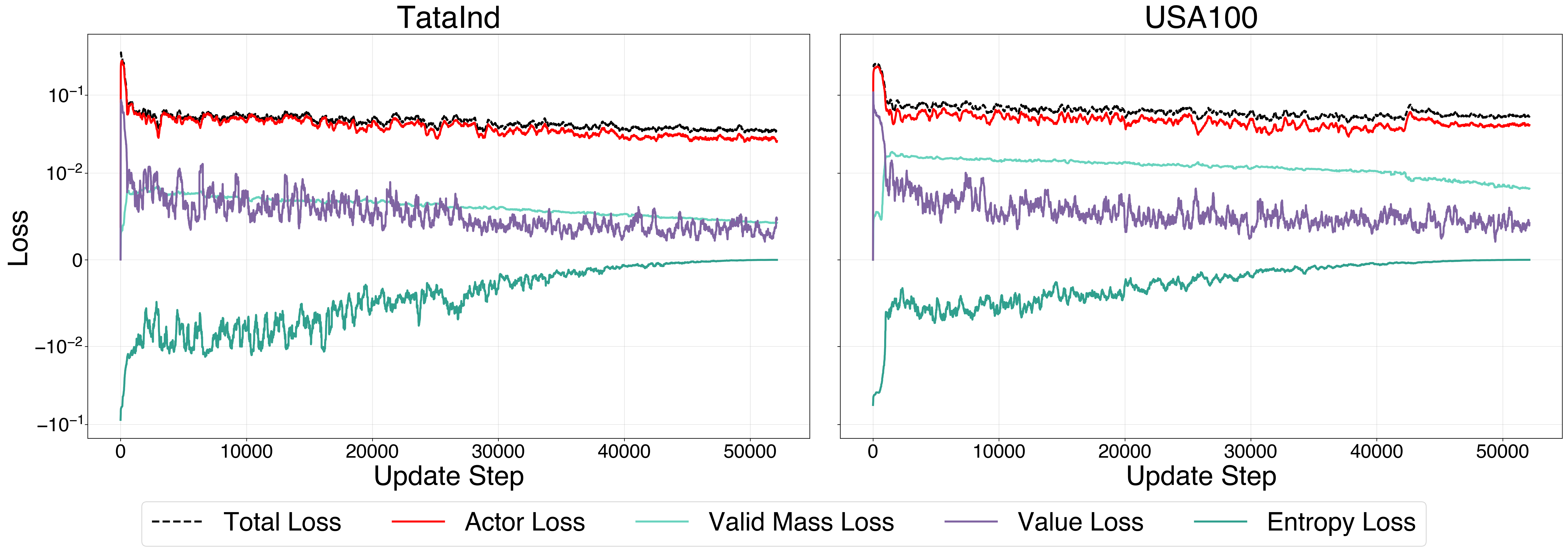}
    \caption{Decomposition of the scalar magnitude of the total training loss into its constituent components (actor, valid mass, value, and entropy losses) over the course of training for TataInd and USA100.}
    \label{fig:loss_components}
\end{figure*}

To provide insight into the optimization process, we decompose the training loss into its constituent components: the actor (policy gradient) loss, valid mass loss, value function loss, and entropy bonus. Figure~\ref{fig:loss_components} shows the signed scalar value of these components over the course of training for both topologies.

The loss dynamics exhibit qualitatively similar behaviour on both topologies. The actor loss dominates the total loss throughout training, showing a steep decrease initially then decreases steadily, reflecting progressive improvement of the policy. 
The valid mass loss starts low (i.e. a moderate amount of the probability mass of the unmasked policy is on valid actions) then increases as the other loss components decrease in magnitude initially, which suggests the share of probability mass on valid actions decreases during this initial learning phase but is counteracted by the valid mass loss. The valid mass component then decreases gradually as $c_{\text{VM}}$ is linearly annealed. The value function loss initially increases from zero when the first blocking events are observed (prior to this the value estimates and observed rewards for successful allocation are all zero, therefore giving zero loss), then gradually decreases over the course of training. The value loss is quite noisy due to the stochastic nature of the generated traffic, which affects the value estimate. The entropy loss (negative, since it acts as a bonus) decreases in magnitude over training as the cosine entropy schedule reduces the entropy coefficient toward zero, allowing the policy to become more deterministic as it converges. The smooth trends in each component, even with a highly stochastic underlying environment, indicate stable optimization throughout the 40M-step training run on both topologies.

\subsection{Blocking Performance}
\label{subsec:blocking_performance}

\begin{figure}[ht]
    \centering
    \includegraphics[width=\linewidth]{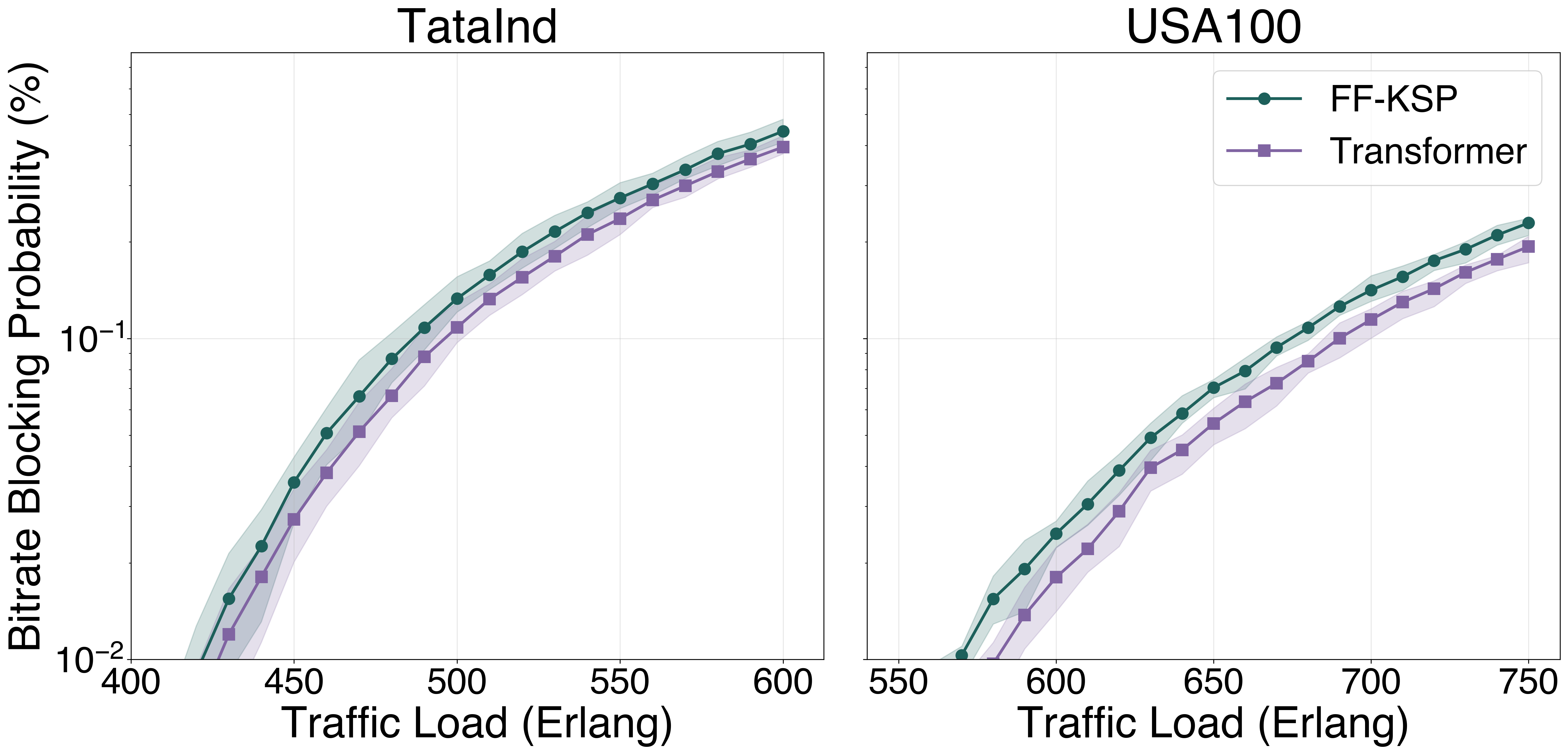}
    \caption{Service blocking probability as a function of traffic load for the Transformer agent and FF-KSP heuristic on TataInd and USA100. Shaded regions indicate the standard error of the mean across parallel environments.}
    \label{fig:blocking_vs_load}
\end{figure}

Figure~\ref{fig:blocking_vs_load} shows the service blocking probability as a function of traffic load for the Graph Transformer agent and the FF-KSP heuristic on both topologies. The Transformer achieves lower blocking than the heuristic across the range of traffic loads. At a target blocking probability of 0.1\%, the Transformer supports approximately 4\% higher traffic load than FF-KSP on USA100 and approximately 3\% higher on TataInd, translating to a meaningful increase in carried traffic on these large-scale networks.

\begin{figure}[ht]
    \centering
    \includegraphics[width=\linewidth]{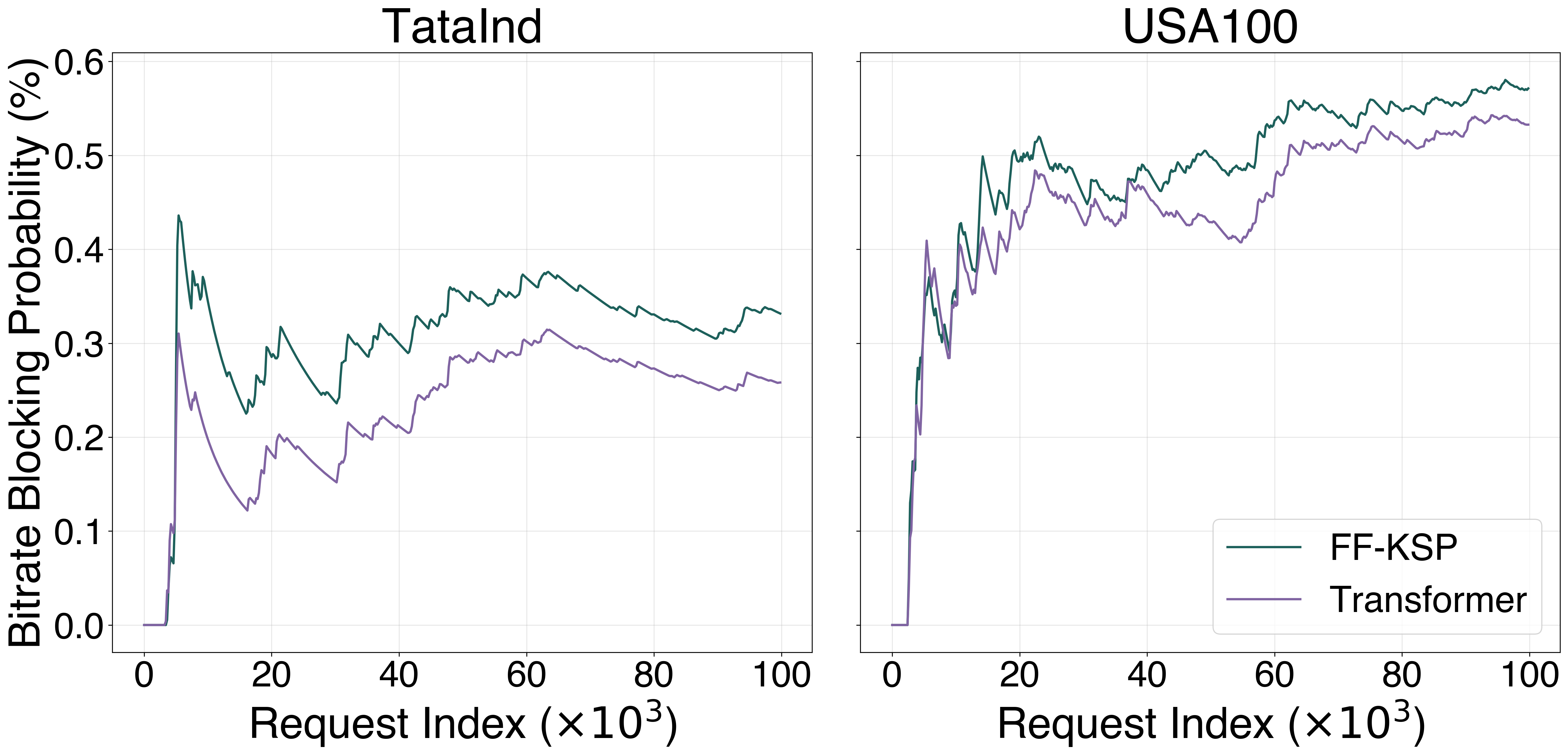}
    \caption{Bitrate blocking probability over the course of a single evaluation episode for the Transformer agent and FF-KSP heuristic on TataInd and USA100.}
    \label{fig:bitrate_blocking_over_steps}
\end{figure}

Figure~\ref{fig:bitrate_blocking_over_steps} shows the bitrate blocking probability over the course of a single evaluation episode, illustrating how blocking evolves as the network transitions from an empty state to steady-state occupancy. During the transient phase (first ${\sim}$10,000 requests), both methods exhibit similar blocking as the network fills. Once steady-state occupancy is reached, the Transformer's advantage emerges and is maintained throughout, indicating a robust learned strategy rather that is insensitive to transient conditions.

% \begin{figure}[htbp]
%     \centering
%     \includegraphics[width=\linewidth]{IMAGES/utilisation_over_steps.png}
%     \caption{Network utilisation over the course of a single evaluation episode for the Transformer agent and FF-KSP heuristic on TataInd and USA100.}
%     \label{fig:utilisation_over_steps}
% \end{figure}

\subsection{Path Selection Analysis}
\label{subsec:path_analysis}

\begin{figure}[ht]
    \centering
    \includegraphics[width=\linewidth]{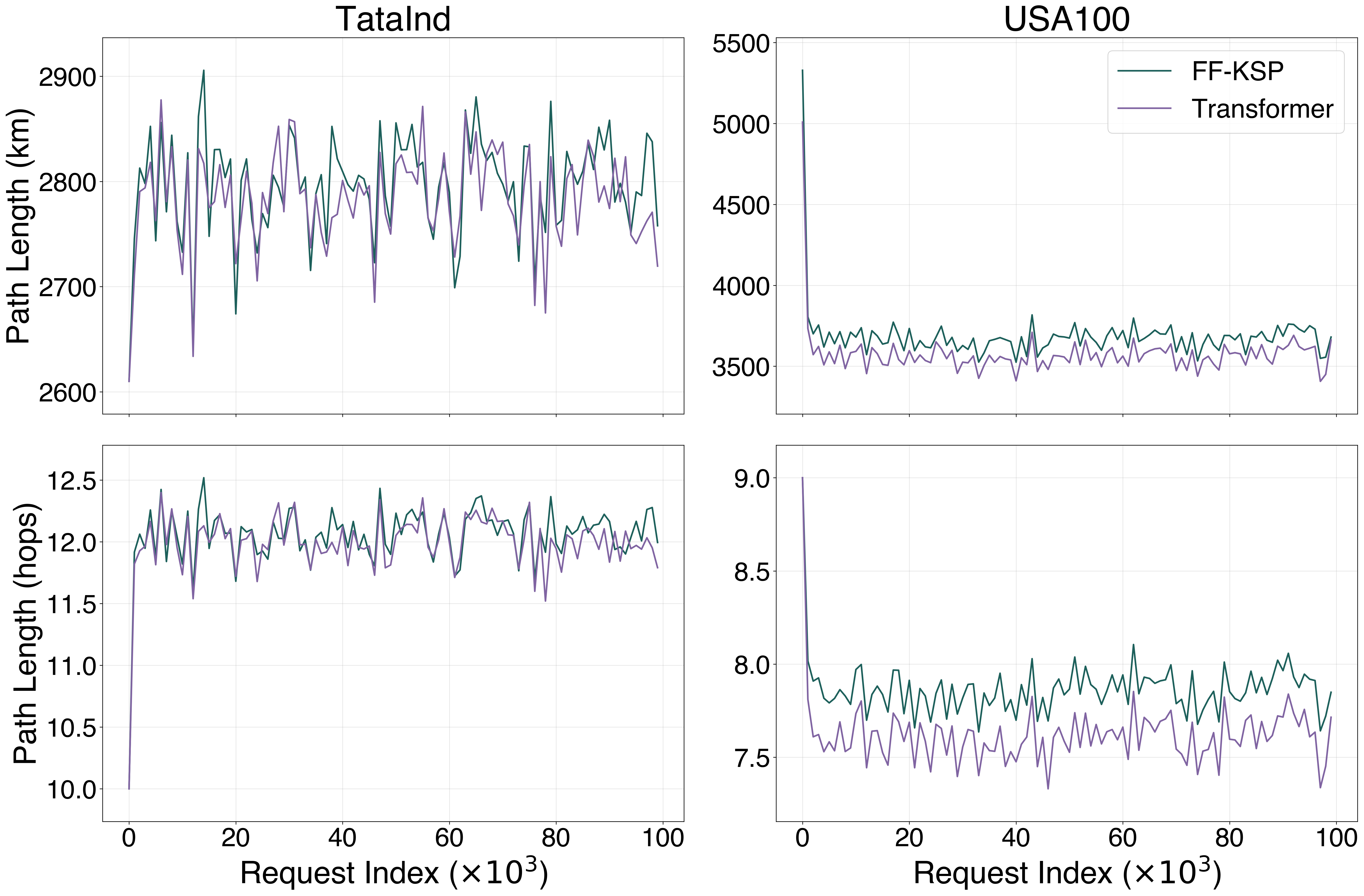}
    \caption{Mean path length in km (top) and hops (bottom) over the course of a single evaluation episode for the Transformer agent and FF-KSP heuristic on TataInd and USA100.}
    \label{fig:path_comparison}
\end{figure}

\begin{figure}[ht]
    \centering
    \includegraphics[width=\linewidth]{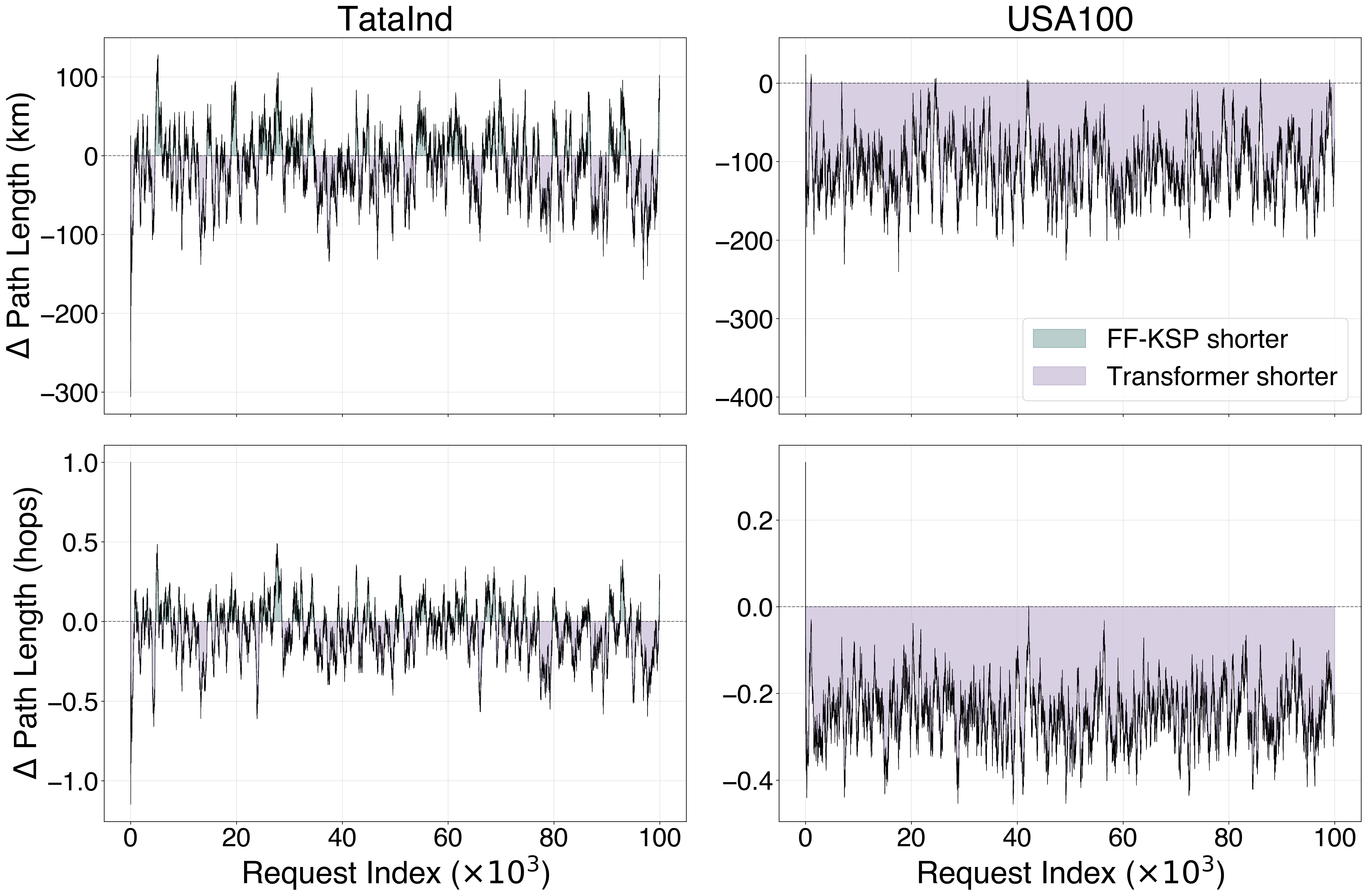}
    \caption{Difference in assigned path length (Transformer minus FF-KSP) in km (top) and hops (bottom) per traffic request for TataInd and USA100. Shaded regions indicate where one method selects shorter paths.}
    \label{fig:path_delta}
\end{figure}

To understand how the Transformer achieves lower blocking, we analyse the routing decisions made by the Transformer and FF-KSP heuristic during evaluation on identical request sets. Figure~\ref{fig:path_comparison} shows the mean path length (in km and hops) over the course of an episode, and Figure~\ref{fig:path_delta} shows the per-request difference in assigned path length, revealing on a request-by-request basis when each method selects shorter or longer paths.

On USA100, the Transformer consistently selects shorter paths than FF-KSP, with a lower mean path length in both km and hops throughout the episode (Figure~\ref{fig:path_comparison}). The per-request delta (Figure~\ref{fig:path_delta}) confirms this is not driven by occasional outliers: the difference is predominantly negative, with a consistent offset of approximately $-$100 to $-$200\,km and $-$0.2 to $-$0.3 hops across the full episode. On TataInd, the mean path lengths are very similar between the two methods, and the per-request delta fluctuates symmetrically around zero, indicating that the Transformer makes different per-request routing decisions without systematically preferring shorter or longer paths. On both topologies, the delta does not change appreciably as the network fills, suggesting a learned structural preference rather than a reactive congestion-avoidance behaviour.

\begin{figure}[ht]
    \centering
    \includegraphics[width=\linewidth]{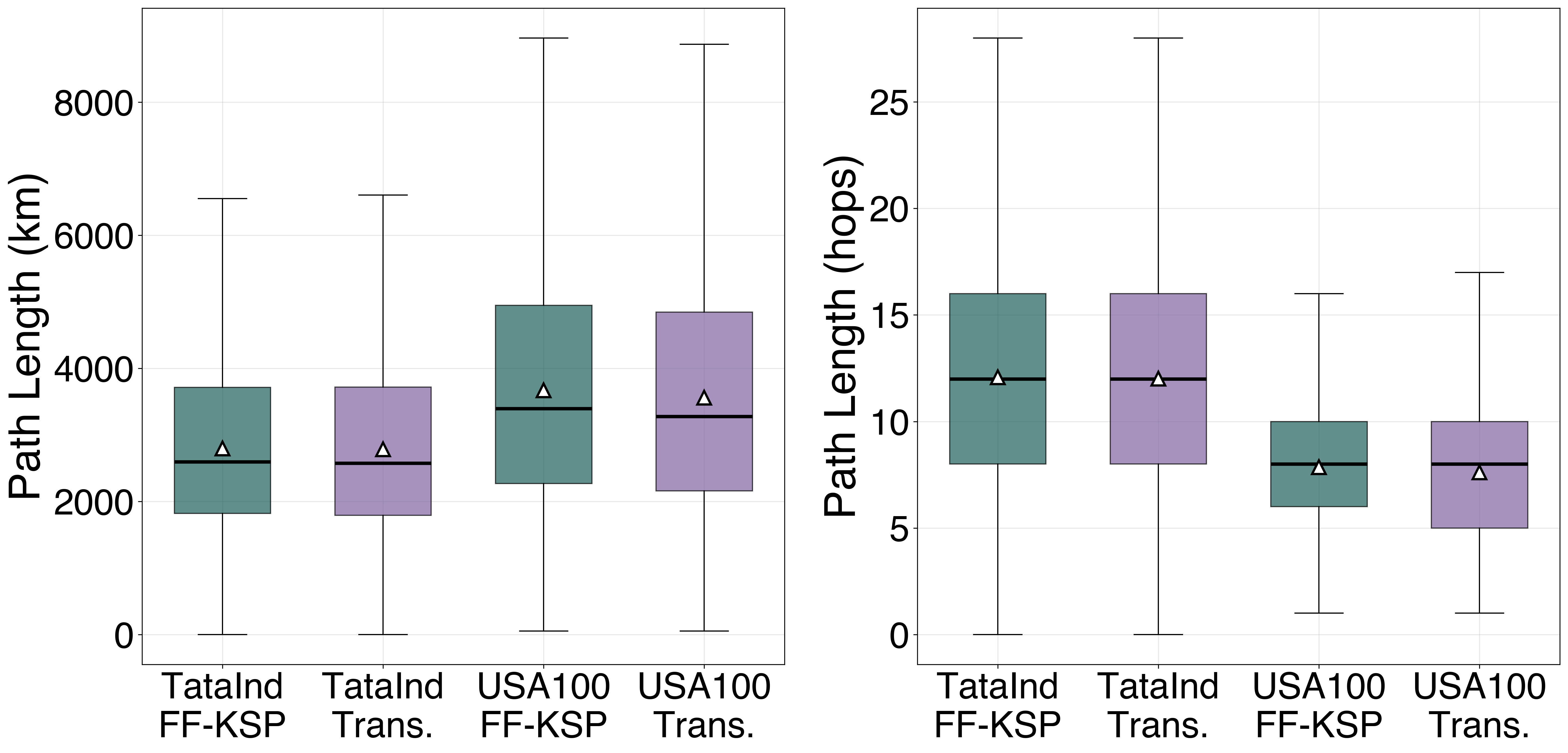}
    \caption{Distribution of assigned path lengths in km (left) and hops (right) for FF-KSP and Transformer across TataInd and USA100. Triangles denote the mean.}
    \label{fig:path_boxplots}
\end{figure}

% TODO - 
Figure~\ref{fig:path_boxplots} presents the full distribution of assigned path lengths for both methods. The Transformer's mean path length on USA100 is approximately 200\,km shorter than FF-KSP, with a correspondingly lower mean hop count. On TataInd, the distributions are nearly identical. Shorter paths enable higher-order modulation formats, consuming fewer FSU per channel and leaving more spectrum available for future requests. This suggests that the Transformer has learned to favour spectrally efficient routes where possible, similar to KSP-FF, while maintaining route diversity like FF-KSP.

\subsection{Spectral Resource Analysis}
\label{subsec:spectral_analysis}

\begin{figure}[ht]
    \centering
    \includegraphics[width=\linewidth]{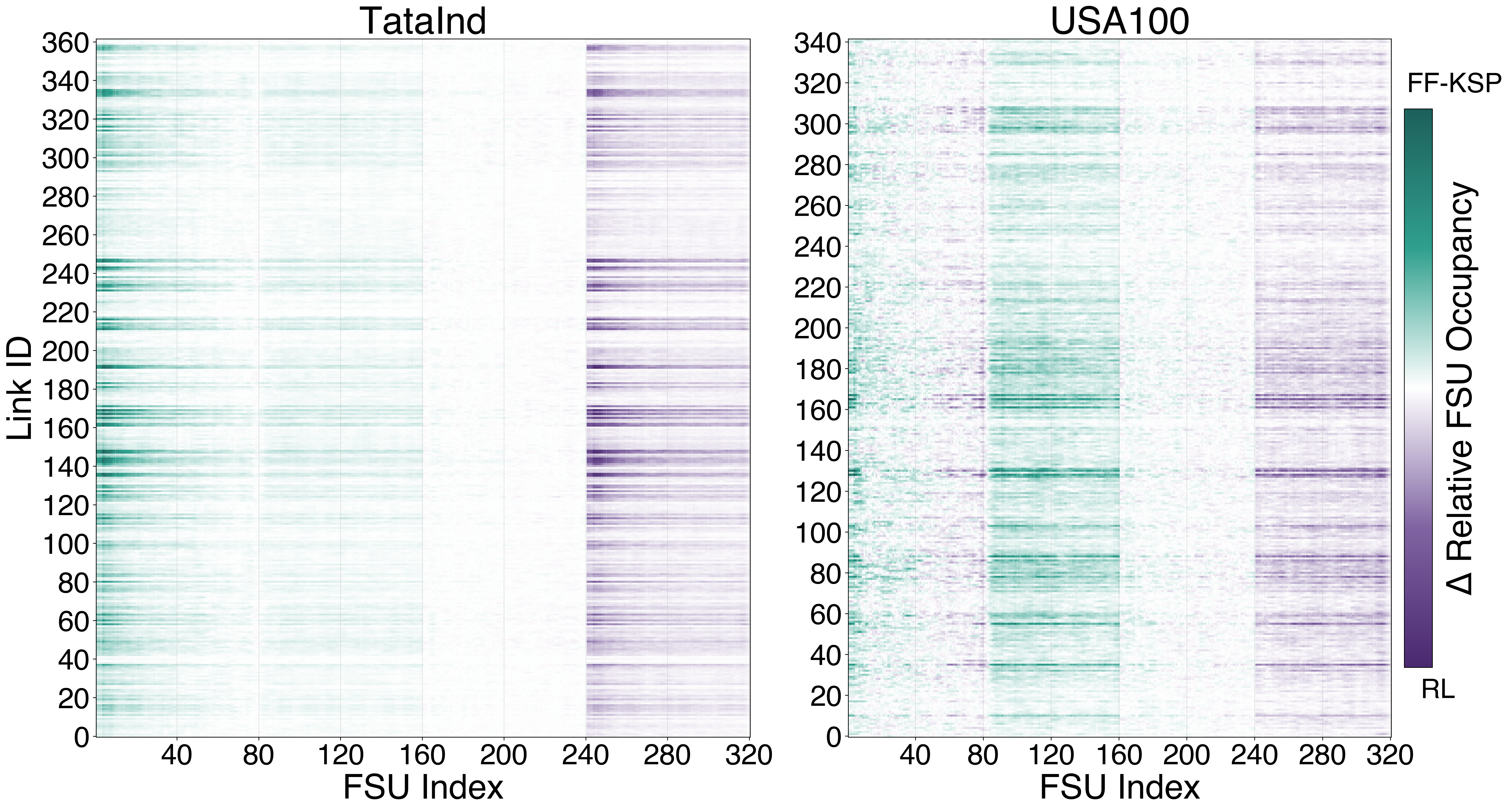}
    \caption{Difference in frequency slot unit (FSU) occupancy between FF-KSP and Transformer across all links for TataInd and USA100. Green indicates higher occupancy by FF-KSP; purple indicates higher occupancy by the Transformer.}
    \label{fig:slot_occupancy_diff}
\end{figure}

\begin{figure}[ht]
    \centering
    \includegraphics[width=\linewidth]{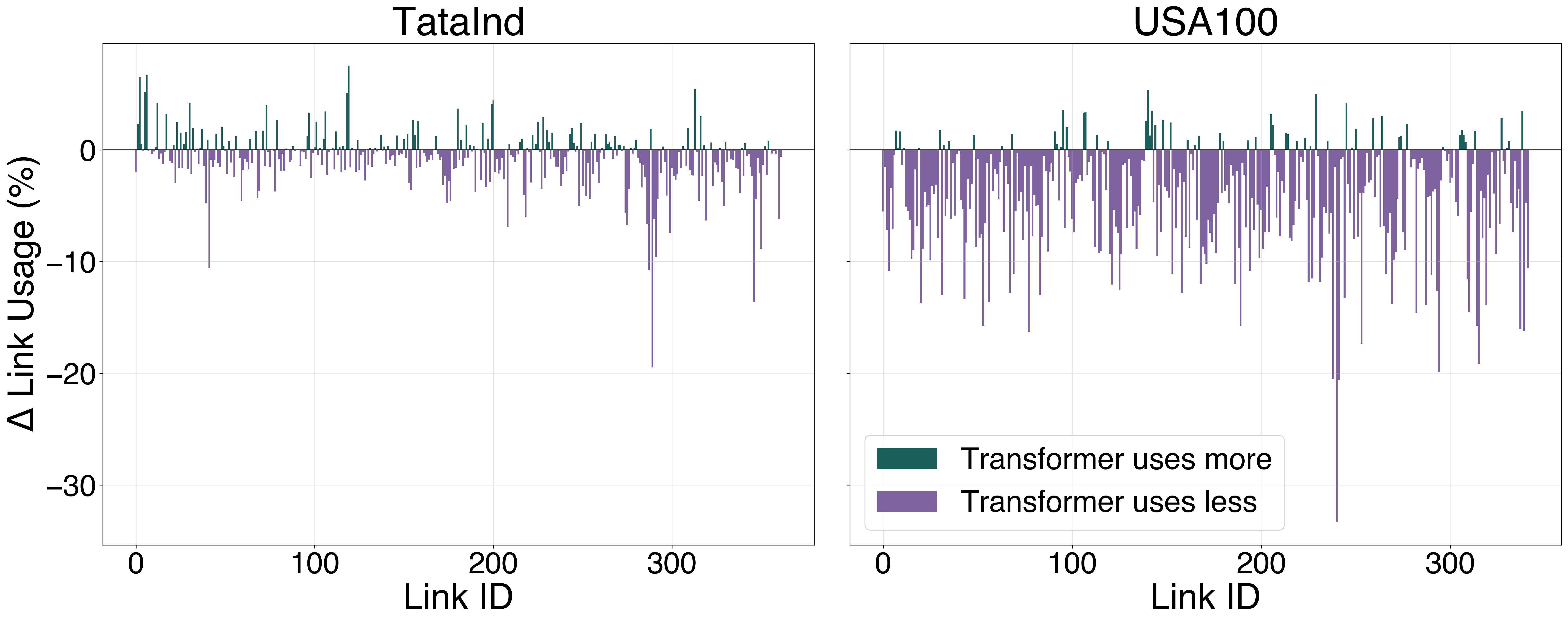}
    \caption{Difference in link usage between Transformer and FF-KSP for each link in TataInd and USA100. Positive values (green) indicate links that have more requests that use them allocated by the Transformer; negative values (purple) indicate links used less.}
    \label{fig:link_usage_delta}
\end{figure}

Figure~\ref{fig:slot_occupancy_diff} shows the difference in FSU occupancy (the percentage of occupied FSU out of total FSU) across all links between the two methods, revealing how the Transformer distributes load differently across the network spectrum. Figure~\ref{fig:link_usage_delta} shows the per-link difference in usage (number of requests that use link), identifying which links the Transformer uses more or less frequently than FF-KSP.

The slot occupancy heatmaps (Figure~\ref{fig:slot_occupancy_diff}) reveal a clear spectral redistribution pattern, particularly on USA100. FF-KSP concentrates occupancy at low FSU indices due to first-fit assignment, while the Transformer distributes allocations more evenly across the spectrum. On USA100, it appears that either end of the spectrum is populated more densely than the centre compared to the heuristic. On TataInd, the same pattern is visible but less pronounced. It may be that further development of this strategy on TataInd could lead to an improved performance gap between the two methods on this topology, closer to the relative performance gap on USA100.

The per-link usage analysis (Figure~\ref{fig:link_usage_delta}) shows that the Transformer uses most links \emph{less} than FF-KSP, with the reduction particularly large on USA100 (up to 20--30\% on some links). This is consistent with the Transformer's preference for shorter paths: fewer hops per lightpath means fewer total link traversals across the network. On TataInd, the link usage differences are smaller and more balanced, with a subset of links seeing increased usage by the Transformer, suggesting a modest rebalancing of traffic across alternative routes. Together, these results indicate that the Transformer achieves lower blocking through shorter, more spectrally efficient routes and more uniform spectrum utilization, rather than the route diversity strategy employed by FF-KSP.

%% file: PAPER_SECTIONS/5_conclusion.tex
\section{Conclusion}
\label{sec:conclusion}

% Room for further optimisation e.g. longer training, hyperparamter tuning, more parallel environments for larger batch sizes with stable gradient estiamtes, staggered environment resets for episodic training \cite{bharthulwar_staggered_2025}.
% Room to perfect the recipe of how the valid mass target and loss coefficents should be tuned, and overall understanding of hyperparameter impacts for this particular family of problems, which XLRON provides an excellent platform for.

We presented the first stable training of a transformer with reinforcement learning for dynamic RMSA in elastic optical networks, combining GPU-accelerated simulation (XLRON), WiRE graph positional encodings, Pre-LayerNorm, off-policy invalid action masking, and valid mass stabilization. On four standard topologies, our Graph Transformer surpasses all previous RL methods and optimized heuristics, achieving up to 13\% higher supported traffic load. On large-scale topologies (TataInd, 143 nodes; USA100, 100 nodes) - the largest dynamic RMSA instances to which RL has been applied - our method supports up to 4\% higher traffic load than the optimized FF-KSP heuristic at $<$0.1\% blocking. Analysis of the trained models revealed distinct routing and spectrum management strategies that provide insight to how learned policies achieve lower blocking, through a combination of short-but-diverse routes and more distributed spectrum allocation than first-fit.

Several future research directions remain. Training could be further improved with greater scaling of batch sizes, other techniques from the RL literature such as staggered environment resets \cite{bharthulwar_staggered_2025}, and the interaction between valid mass target, loss coefficients, and other hyperparameters warrants further investigation. Since our method is able to learn superior policies for a single objective such as blocking probability minimization, it is natural to extend it to a multi-objective setting \cite{nallaperuma_interpreting_2023} or joint optimization of parameters such as launch power, where we expect its advantages to be even more pronounced.

Having established that our method solves R(M)SA under distance-dependent physical layer assumptions, a natural next step is to incorporate more realistic models - including analytical SNR estimation with nonlinear interference - to determine whether the throughput advantage of our method over heuristics can be improved with QoT-aware allocation. Since our architecture can flexibly process any per-link data as a token and we have shown it scales to large problem instances, it can be applied to multi-band or multi-core networks with per-channel SNR values on links.

All code and experimental data are openly available at \cite{doherty_xlron_2023} to support reproducibility and future benchmarking.

%% file: references.bib
@article{teng_drl-assisted_2025,
	title = {{DRL}-{Assisted} {QoT}-{Aware} {Service} {Provisioning} in {Multi}-{Band} {Elastic} {Optical} {Networks}},
	volume = {43},
	copyright = {https://ieeexplore.ieee.org/Xplorehelp/downloads/license-information/IEEE.html},
	issn = {0733-8724, 1558-2213},
	url = {https://ieeexplore.ieee.org/document/11131684/},
	doi = {10.1109/JLT.2025.3601402},
	number = {19},
	urldate = {2026-03-27},
	journal = {Journal of Lightwave Technology},
	author = {Teng, Yiran and Natalino, Carlos and Arpanaei, Farhad and Li, Haiyuan and SÃ¡nchez-MaciÃ¡n, Alfonso and Monti, Paolo and Yan, Shuangyi and Simeonidou, Dimitra},
	month = oct,
	year = {2025},
	pages = {9090--9101},
}

@misc{su_roformer_2023,
	title = {{RoFormer}: {Enhanced} {Transformer} with {Rotary} {Position} {Embedding}},
	shorttitle = {{RoFormer}},
	url = {http://arxiv.org/abs/2104.09864},
	doi = {10.48550/arXiv.2104.09864},
	urldate = {2026-03-24},
	publisher = {arXiv},
	author = {Su, Jianlin and Lu, Yu and Pan, Shengfeng and Murtadha, Ahmed and Wen, Bo and Liu, Yunfeng},
	month = nov,
	year = {2023},
	note = {arXiv:2104.09864 [cs]},
}

@misc{vaswani_attention_2017,
	title = {Attention {Is} {All} {You} {Need}},
	copyright = {arXiv.org perpetual, non-exclusive license},
	url = {https://arxiv.org/abs/1706.03762},
	doi = {10.48550/ARXIV.1706.03762},
	urldate = {2026-03-24},
	publisher = {arXiv},
	author = {Vaswani, Ashish and Shazeer, Noam and Parmar, Niki and Uszkoreit, Jakob and Jones, Llion and Gomez, Aidan N. and Kaiser, Lukasz and Polosukhin, Illia},
	year = {2017},
	note = {Version Number: 7},
}

@article{hou_exploring_2023,
	title = {Exploring the {Use} of {Invalid} {Action} {Masking} in {Reinforcement} {Learning}: {A} {Comparative} {Study} of {On}-{Policy} and {Off}-{Policy} {Algorithms} in {Real}-{Time} {Strategy} {Games}},
	volume = {13},
	issn = {2076-3417},
	shorttitle = {Exploring the {Use} of {Invalid} {Action} {Masking} in {Reinforcement} {Learning}},
	url = {https://www.mdpi.com/2076-3417/13/14/8283},
	doi = {10.3390/app13148283},
	language = {en},
	number = {14},
	urldate = {2026-03-24},
	journal = {Applied Sciences},
	author = {Hou, Yueqi and Liang, Xiaolong and Zhang, Jiaqiang and Yang, Qisong and Yang, Aiwu and Wang, Ning},
	month = jul,
	year = {2023},
	pages = {8283},
}

@article{chen_transformer-pointer_2026,
	title = {Transformer-pointer {DRL} model for static resource allocation problems in {SDM}-{EONs}},
	volume = {18},
	issn = {1943-0620, 1943-0639},
	url = {https://opg.optica.org/abstract.cfm?URI=jocn-18-3-315},
	doi = {10.1364/JOCN.580228},
	language = {en},
	number = {3},
	urldate = {2026-03-05},
	journal = {Journal of Optical Communications and Networking},
	author = {Chen, Sibo and Wang, Jiading and Shigeno, Maiko},
	month = mar,
	year = {2026},
	pages = {315},
}

@misc{brody_how_2022,
	title = {How {Attentive} are {Graph} {Attention} {Networks}?},
	url = {http://arxiv.org/abs/2105.14491},
	doi = {10.48550/arXiv.2105.14491},
	urldate = {2026-03-24},
	publisher = {arXiv},
	author = {Brody, Shaked and Alon, Uri and Yahav, Eran},
	month = jan,
	year = {2022},
	note = {arXiv:2105.14491 [cs]},
}

@misc{velickovic_graph_2018,
	title = {Graph {Attention} {Networks}},
	url = {http://arxiv.org/abs/1710.10903},
	doi = {10.48550/arXiv.1710.10903},
	urldate = {2026-03-24},
	publisher = {arXiv},
	author = {VeliÄkoviÄ, Petar and Cucurull, Guillem and Casanova, Arantxa and Romero, Adriana and LiÃ², Pietro and Bengio, Yoshua},
	month = feb,
	year = {2018},
	note = {arXiv:1710.10903 [stat]},
}

@misc{reid_rotary_2026,
	title = {Rotary {Position} {Encodings} for {Graphs}},
	url = {http://arxiv.org/abs/2509.22259},
	doi = {10.48550/arXiv.2509.22259},
	urldate = {2026-03-24},
	publisher = {arXiv},
	author = {Reid, Isaac and Sehanobish, Arijit and HÃ¶fs, Cederik and Mlodozeniec, Bruno and Vulpius, Leonhard and Barbero, Federico and Weller, Adrian and Choromanski, Krzysztof and Turner, Richard E. and VeliÄkoviÄ, Petar},
	month = jan,
	year = {2026},
	note = {arXiv:2509.22259 [cs]},
}

@misc{black_comparing_2024,
	title = {Comparing {Graph} {Transformers} via {Positional} {Encodings}},
	url = {http://arxiv.org/abs/2402.14202},
	doi = {10.48550/arXiv.2402.14202},
	urldate = {2026-03-24},
	publisher = {arXiv},
	author = {Black, Mitchell and Wan, Zhengchao and Mishne, Gal and Nayyeri, Amir and Wang, Yusu},
	month = aug,
	year = {2024},
	note = {arXiv:2402.14202 [cs]},
}

@misc{ma_graph_2023,
	title = {Graph {Inductive} {Biases} in {Transformers} without {Message} {Passing}},
	url = {http://arxiv.org/abs/2305.17589},
	doi = {10.48550/arXiv.2305.17589},
	urldate = {2026-03-24},
	publisher = {arXiv},
	author = {Ma, Liheng and Lin, Chen and Lim, Derek and Romero-Soriano, Adriana and Dokania, Puneet K. and Coates, Mark and Torr, Philip and Lim, Ser-Nam},
	month = may,
	year = {2023},
	note = {arXiv:2305.17589 [cs]},
}

@misc{ying_transformers_2021,
	title = {Do {Transformers} {Really} {Perform} {Bad} for {Graph} {Representation}?},
	url = {http://arxiv.org/abs/2106.05234},
	doi = {10.48550/arXiv.2106.05234},
	urldate = {2026-03-24},
	publisher = {arXiv},
	author = {Ying, Chengxuan and Cai, Tianle and Luo, Shengjie and Zheng, Shuxin and Ke, Guolin and He, Di and Shen, Yanming and Liu, Tie-Yan},
	month = nov,
	year = {2021},
	note = {arXiv:2106.05234 [cs]},
}

@misc{parisotto_stabilizing_2019,
	title = {Stabilizing {Transformers} for {Reinforcement} {Learning}},
	url = {http://arxiv.org/abs/1910.06764},
	doi = {10.48550/arXiv.1910.06764},
	urldate = {2026-03-24},
	publisher = {arXiv},
	author = {Parisotto, Emilio and Song, H. Francis and Rae, Jack W. and Pascanu, Razvan and Gulcehre, Caglar and Jayakumar, Siddhant M. and Jaderberg, Max and Kaufman, Raphael Lopez and Clark, Aidan and Noury, Seb and Botvinick, Matthew M. and Heess, Nicolas and Hadsell, Raia},
	month = oct,
	year = {2019},
	note = {arXiv:1910.06764 [cs]},
}

@misc{chen_decision_2021,
	title = {Decision {Transformer}: {Reinforcement} {Learning} via {Sequence} {Modeling}},
	shorttitle = {Decision {Transformer}},
	url = {http://arxiv.org/abs/2106.01345},
	doi = {10.48550/arXiv.2106.01345},
	urldate = {2026-03-24},
	publisher = {arXiv},
	author = {Chen, Lili and Lu, Kevin and Rajeswaran, Aravind and Lee, Kimin and Grover, Aditya and Laskin, Michael and Abbeel, Pieter and Srinivas, Aravind and Mordatch, Igor},
	month = jun,
	year = {2021},
	note = {arXiv:2106.01345 [cs]},
}

@misc{xiong_layer_2020,
	title = {On {Layer} {Normalization} in the {Transformer} {Architecture}},
	url = {http://arxiv.org/abs/2002.04745},
	doi = {10.48550/arXiv.2002.04745},
	urldate = {2026-03-24},
	publisher = {arXiv},
	author = {Xiong, Ruibin and Yang, Yunchang and He, Di and Zheng, Kai and Zheng, Shuxin and Xing, Chen and Zhang, Huishuai and Lan, Yanyan and Wang, Liwei and Liu, Tie-Yan},
	month = jun,
	year = {2020},
	note = {arXiv:2002.04745 [cs]},
}

@article{wang_physical_2026,
	title = {Physical layer-aware deep reinforcement learning with advantage function stabilization for dynamic {RMSA} in elastic optical networks},
	volume = {18},
	issn = {1943-0620, 1943-0639},
	url = {https://opg.optica.org/abstract.cfm?URI=jocn-18-3-250},
	doi = {10.1364/JOCN.577029},
	language = {en},
	number = {3},
	urldate = {2026-02-23},
	journal = {Journal of Optical Communications and Networking},
	author = {Wang, Haojie and Wang, Yixin and Zhao, Yongli and Zhang, Jie},
	month = mar,
	year = {2026},
	pages = {250},
}

@article{doherty_reinforcement_2025,
	title = {Reinforcement learning for dynamic resource allocation in optical networks: hype or hope?},
	volume = {17},
	issn = {1943-0620, 1943-0639},
	shorttitle = {Reinforcement learning for dynamic resource allocation in optical networks},
	url = {https://opg.optica.org/abstract.cfm?URI=jocn-17-9-D1},
	doi = {10.1364/JOCN.559990},
	language = {en},
	number = {9},
	urldate = {2026-02-11},
	journal = {Journal of Optical Communications and Networking},
	author = {Doherty, Michael and Matzner, Robin and Sadeghi, Rasoul and Bayvel, Polina and Beghelli, Alejandra},
	month = sep,
	year = {2025},
	pages = {D1},
}

@misc{bharthulwar_staggered_2025,
	title = {Staggered {Environment} {Resets} {Improve} {Massively} {Parallel} {On}-{Policy} {Reinforcement} {Learning}},
	url = {http://arxiv.org/abs/2511.21011},
	doi = {10.48550/arXiv.2511.21011},
	urldate = {2025-12-16},
	publisher = {arXiv},
	author = {Bharthulwar, Sid and Tao, Stone and Su, Hao},
	month = nov,
	year = {2025},
	note = {arXiv:2511.21011 [cs]},
}

@misc{naik_reward_2024,
	title = {Reward {Centering}},
	url = {http://arxiv.org/abs/2405.09999},
	doi = {10.48550/arXiv.2405.09999},
	urldate = {2025-04-03},
	publisher = {arXiv},
	author = {Naik, Abhishek and Wan, Yi and Tomar, Manan and Sutton, Richard S.},
	month = oct,
	year = {2024},
	note = {arXiv:2405.09999 [cs]},
}

@inproceedings{xiong_graph_2024,
	address = {Beijing, China},
	title = {Graph {Attention} {Network} {Enhanced} {Deep} {Reinforcement} {Learning} {Framework} for {Routing}, {Modulation}, and {Spectrum} {Allocation} in {EONs}},
	copyright = {https://doi.org/10.15223/policy-029},
	isbn = {9798350379266},
	url = {https://ieeexplore.ieee.org/document/10810116/},
	doi = {10.1109/ACP/IPOC63121.2024.10810116},
	urldate = {2025-01-03},
	booktitle = {2024 {Asia} {Communications} and {Photonics} {Conference} ({ACP}) and {International} {Conference} on {Information} {Photonics} and {Optical} {Communications} ({IPOC})},
	publisher = {IEEE},
	author = {Xiong, Zheng and Huang, Yue-Cai and Hu, Xiaohui},
	month = nov,
	year = {2024},
	pages = {1--6},
}

@inproceedings{jaumard_decomposition_2023,
	title = {Decomposition {Models} for the {Routing} and {Slot} {Provisioning} {Problem}},
	doi = {10.1109/ICNC57223.2023.10074088},
	booktitle = {2023 {International} {Conference} on {Computing}, {Networking} and {Communications} ({ICNC})},
	author = {Jaumard, Brigitte and Mohammed, Adham and Nguyen, Quang Anh},
	year = {2023},
	pages = {659--665},
}

@inproceedings{shimoda_mask_2021,
	address = {Bordeaux, France},
	title = {Mask {RSA}: {End}-{To}-{End} {Reinforcement} {Learning}-based {Routing} and {Spectrum} {Assignment} in {Elastic} {Optical} {Networks}},
	isbn = {978-1-66543-868-1},
	shorttitle = {Mask {RSA}},
	url = {https://ieeexplore.ieee.org/document/9606169/},
	doi = {10.1109/ECOC52684.2021.9606169},
	language = {en},
	urldate = {2023-01-09},
	booktitle = {2021 {European} {Conference} on {Optical} {Communication} ({ECOC})},
	publisher = {IEEE},
	author = {Shimoda, Masayuki and Tanaka, Takafumi},
	month = sep,
	year = {2021},
	pages = {1--4},
}

@phdthesis{beghelli_resource_2006,
	type = {Ph.{D}. thesis},
	title = {Resource allocation and scalability in dynamic wavelength-routed optical networks},
	url = {https://discovery.ucl.ac.uk/id/eprint/1445164},
	school = {University of London},
	author = {Beghelli, A.L.Z.},
	year = {2006},
}

@phdthesis{baroni_routing_1998,
	address = {United Kingdom},
	type = {Ph.{D}. thesis},
	title = {Routing and wavelength allocation in {WDM} optical networks},
	url = {https://discovery.ucl.ac.uk/id/eprint/10100693},
	school = {University College London},
	author = {Baroni, Stefano},
	year = {1998},
}

@misc{matzner_topology_2024,
	title = {Topology {Bench}: {Systematic} {Graph} {Based} {Benchmarking} for {Core} {Optical} {Networks}},
	copyright = {Creative Commons Attribution 4.0 International},
	shorttitle = {Topology {Bench}},
	url = {https://arxiv.org/abs/2411.04160},
	doi = {10.48550/ARXIV.2411.04160},
	urldate = {2024-11-29},
	publisher = {arXiv},
	author = {Matzner, Robin and Ahuja, Akanksha and Sadeghi, Rasoul and Doherty, Michael and Beghelli, Alejandra and Savory, Seb J. and Bayvel, Polina},
	year = {2024},
	note = {Version Number: 1},
}

@inproceedings{doherty_xlron_2024,
	title = {{XLRON}: {Accelerated} {Reinforcement} {Learning} {Environments} for {Optical} {Networks}},
	booktitle = {2024 {Optical} {Fiber} {Communications} {Conference} and {Exhibition} ({OFC})},
	author = {Doherty, Michael and Beghelli, Alejandra},
	year = {2024},
	pages = {1--3},
}

@article{teng_deep-reinforcement-learning-based_2024,
	title = {Deep-reinforcement-learning-based {RMSCA} for space division multiplexing networks with multi-core fibers [{Invited} {Tutorial}]},
	volume = {16},
	issn = {1943-0620, 1943-0639},
	url = {https://opg.optica.org/abstract.cfm?URI=jocn-16-7-C76},
	doi = {10.1364/JOCN.518685},
	language = {en},
	number = {7},
	urldate = {2024-08-12},
	journal = {Journal of Optical Communications and Networking},
	author = {Teng, Yiran and Natalino, Carlos and Li, Haiyuan and Yang, Ruizhi and Majeed, Jassim and Shen, Sen and Monti, Paolo and Nejabati, Reza and Yan, Shuangyi and Simeonidou, Dimitra},
	month = jul,
	year = {2024},
	pages = {C76},
}

@article{cheng_ptrnet-rsa_2024,
	title = {{PtrNet}-{RSA}: {A} {Pointer} {Network}-based {QoT}-aware {Routing} and {Spectrum} {Assignment} {Scheme} in {Elastic} {Optical} {Networks}},
	doi = {10.1109/JLT.2024.3405587},
	journal = {Journal of Lightwave Technology},
	author = {Cheng, Yuansen and Ding, Shifeng and Shao, Yingjie and Chan, Chun-Kit},
	year = {2024},
	pages = {1--12},
}

@article{nallaperuma_interpreting_2023,
	title = {Interpreting multi-objective reinforcement learning for routing and wavelength assignment in optical networks},
	volume = {15},
	issn = {1943-0620, 1943-0639},
	url = {https://opg.optica.org/abstract.cfm?URI=jocn-15-8-497},
	doi = {10.1364/JOCN.483733},
	language = {en},
	number = {8},
	urldate = {2023-07-25},
	journal = {Journal of Optical Communications and Networking},
	author = {Nallaperuma, Sam and Gan, Zelin and Nevin, Josh and Shevchenko, Mykyta and Savory, Seb J.},
	month = aug,
	year = {2023},
	pages = {497},
}

@article{tang_heuristic_2022,
	title = {Heuristic {Reward} {Design} for {Deep} {Reinforcement} {Learning}-{Based} {Routing}, {Modulation} and {Spectrum} {Assignment} of {Elastic} {Optical} {Networks}},
	volume = {26},
	issn = {1089-7798, 1558-2558, 2373-7891},
	url = {https://ieeexplore.ieee.org/document/9847208/},
	doi = {10.1109/LCOMM.2022.3195778},
	number = {11},
	urldate = {2023-07-08},
	journal = {IEEE Communications Letters},
	author = {Tang, Bixia and Huang, Yue-Cai and Xue, Yun and Zhou, Weixing},
	month = nov,
	year = {2022},
	pages = {2675--2679},
}

@article{xu_deep_2022,
	title = {Deep {Reinforcement} {Learning}-{Based} {Routing} and {Spectrum} {Assignment} of {EONs} by {Exploiting} {GCN} and {RNN} for {Feature} {Extraction}},
	volume = {40},
	doi = {10.1109/JLT.2022.3175865},
	number = {15},
	journal = {Journal of Lightwave Technology},
	author = {Xu, Liufei and Huang, Yue-Cai and Xue, Yun and Hu, Xiaohui},
	year = {2022},
	pages = {4945--4955},
}

@article{chen_deeprmsa_2019,
	title = {{DeepRMSA}: {A} {Deep} {Reinforcement} {Learning} {Framework} for {Routing}, {Modulation} and {Spectrum} {Assignment} in {Elastic} {Optical} {Networks}},
	volume = {37},
	issn = {0733-8724, 1558-2213},
	shorttitle = {{DeepRMSA}},
	url = {https://ieeexplore.ieee.org/document/8738827/},
	doi = {10.1109/JLT.2019.2923615},
	language = {en},
	number = {16},
	urldate = {2023-01-09},
	journal = {Journal of Lightwave Technology},
	author = {Chen, Xiaoliang and Li, Baojia and Proietti, Roberto and Lu, Hongbo and Zhu, Zuqing and Yoo, S. J. Ben},
	month = aug,
	year = {2019},
	pages = {4155--4163},
}

@inproceedings{garcia_multicast_2003,
	address = {Zagreb, Croatia},
	title = {A multicast reinforcement learning algorithm for {WDM} optical networks},
	isbn = {978-953-184-052-1},
	url = {http://ieeexplore.ieee.org/document/1215852/},
	doi = {10.1109/CONTEL.2003.176942},
	urldate = {2024-08-12},
	booktitle = {Proceedings of the 7th {International} {Conference} on {Telecommunications}, 2003. {ConTEL} 2003.},
	publisher = {IEEE},
	author = {Garcia, P. and Zsigri, A. and Guitton, A.},
	year = {2003},
	pages = {419--426 vol.2},
}

@inproceedings{cruzado_capacity-bound_2024,
	title = {Capacity-{Bound} {Evaluation} and {Routing} and {Spectrum} {Assignment} for {Elastic} {Optical} {Path} {Networks} with {Distance}-{Adaptive} {Modulation}},
	booktitle = {2024 {Optical} {Fiber} {Communications} {Conference} and {Exhibition} ({OFC})},
	author = {Cruzado, Kenji and Mori, Yojiro and Lin, Shih-Chun and Matsuura, Motoharu and Subramaniam, Suresh and Hasegawa, Hiroshi},
	year = {2024},
	pages = {1--3},
}

@misc{vinyals_pointer_2015,
	title = {Pointer {Networks}},
	copyright = {arXiv.org perpetual, non-exclusive license},
	url = {https://arxiv.org/abs/1506.03134},
	doi = {10.48550/ARXIV.1506.03134},
	urldate = {2024-06-30},
	publisher = {arXiv},
	author = {Vinyals, Oriol and Fortunato, Meire and Jaitly, Navdeep},
	year = {2015},
	note = {Version Number: 2},
}

@article{hayashi_cost-effective_2023,
	title = {Cost-effective network capacity upgrade by heterogeneous wavelength division multiplexing density with bandwidth-variable virtual direct links},
	volume = {15},
	doi = {10.1364/JOCN.485116},
	number = {9},
	journal = {Journal of Optical Communications and Networking},
	author = {Hayashi, Kazuki and Mori, Yojiro and Hasegawa, Hiroshi},
	year = {2023},
	pages = {D23--D32},
}

@inproceedings{cruzado_effective_2023,
	title = {Effective {Capacity} {Estimation} {Based} on {Cut}-{Set} {Load} {Analysis} in {Optical} {Path} {Networks}},
	doi = {10.1109/PSC57974.2023.10297165},
	booktitle = {2023 {International} {Conference} on {Photonics} in {Switching} and {Computing} ({PSC})},
	author = {Cruzado, Kenji and Mori, Yojiro and Lin, Shih-Chun and Matsuura, Motoharu and Subramaniam, Suresh and Hasegawa, Hiroshi},
	year = {2023},
	pages = {1--3},
}

@misc{doherty_xlron_2023,
	title = {{XLRON}: {Accelerated} {Learning} and {Resource} {Allocation} for {Optical} {Networks}},
	url = {https://github.com/micdoh/XLRON.git},
	howpublished = {\url{https://github.com/micdoh/XLRON.git}},
	author = {Doherty, Michael},
	month = sep,
	year = {2023},
}

@misc{hessel_podracer_2021,
	title = {Podracer architectures for scalable {Reinforcement} {Learning}},
	url = {http://arxiv.org/abs/2104.06272},
	urldate = {2023-09-20},
	publisher = {arXiv},
	author = {Hessel, Matteo and Kroiss, Manuel and Clark, Aidan and Kemaev, Iurii and Quan, John and Keck, Thomas and Viola, Fabio and van Hasselt, Hado},
	month = apr,
	year = {2021},
	note = {arXiv:2104.06272 [cs]},
}

@misc{huang_action_2020,
	title = {Action {Guidance}: {Getting} the {Best} of {Sparse} {Rewards} and {Shaped} {Rewards} for {Real}-time {Strategy} {Games}},
	shorttitle = {Action {Guidance}},
	url = {http://arxiv.org/abs/2010.03956},
	urldate = {2023-07-08},
	publisher = {arXiv},
	author = {Huang, Shengyi and Ontañón, Santiago},
	month = oct,
	year = {2020},
	note = {arXiv:2010.03956 [cs, stat]},
}

@misc{schulman_high-dimensional_2018,
	title = {High-{Dimensional} {Continuous} {Control} {Using} {Generalized} {Advantage} {Estimation}},
	url = {http://arxiv.org/abs/1506.02438},
	urldate = {2023-07-08},
	publisher = {arXiv},
	author = {Schulman, John and Moritz, Philipp and Levine, Sergey and Jordan, Michael and Abbeel, Pieter},
	month = oct,
	year = {2018},
	note = {arXiv:1506.02438 [cs]},
}

@misc{schulman_proximal_2017,
	title = {Proximal {Policy} {Optimization} {Algorithms}},
	url = {http://arxiv.org/abs/1707.06347},
	doi = {10.48550/arXiv.1707.06347},
	urldate = {2023-01-20},
	publisher = {arXiv},
	author = {Schulman, John and Wolski, Filip and Dhariwal, Prafulla and Radford, Alec and Klimov, Oleg},
	month = aug,
	year = {2017},
	note = {arXiv:1707.06347 [cs]},
}

@misc{ba_layer_2016,
	title = {Layer Normalization},
	url = {http://arxiv.org/abs/1607.06450},
	urldate = {2024-01-15},
	publisher = {arXiv},
	author = {Ba, Jimmy Lei and Kiros, Jamie Ryan and Hinton, Geoffrey E.},
	month = jul,
	year = {2016},
	note = {arXiv:1607.06450 [cs, stat]},
}

@misc{ennadir2025poolmewisely,
	title = {Pool Me Wisely: Rethinking Graph Pooling in Graph Transformers},
	url = {http://arxiv.org/abs/2502.11032},
	author = {Ennadir, Soufiane and Vazirgiannis, Michalis and Liao, Renjie},
	year = {2025},
	note = {arXiv:2502.11032},
}

@misc{zabounidis_overcoming_2026,
	title = {Overcoming {Valid} {Action} {Suppression} in {Unmasked} {Policy} {Gradient} {Algorithms}},
	url = {http://arxiv.org/abs/2603.09090},
	doi = {10.48550/arXiv.2603.09090},
	urldate = {2026-03-24},
	publisher = {arXiv},
	author = {Zabounidis, Renos and Siegelmann, Roy and Qadri, Mohamad and Kim, Woojun and Stepputtis, Simon and Sycara, Katia P.},
	month = mar,
	year = {2026},
	note = {arXiv:2603.09090 [cs]},
}
